\documentclass[twocolumn]{aastex63}

\usepackage{amssymb}	
\usepackage{amsmath}	
\usepackage{array}
\usepackage{bm}
\usepackage{CJK}
\usepackage{graphicx}	
\usepackage{hyperref}
\usepackage{xifthen}
\usepackage{xspace}
\usepackage{natbib}

\begin{document}
\begin{CJK*}{UTF8}{gbsn}
\title{Vortex-Induced Rings and Gaps within Protoplanetary Disks}

\author[0000-0003-2622-6895]{Xiaoyi Ma (马潇依)}
\affiliation{Kavli Institute for Astronomy and Astrophysics, Peking University, Beijing 100871, China}
\affiliation{Department of Astronomy, School of Physics, Peking University, Beijing 100871, China}

\author{Pinghui Huang (黄平辉)}
\affiliation{Department of Physics and Astronomy, University of Victoria, Victoria, BC, V8P 5C2, Canada}

\author[0000-0003-0454-7890]{Cong Yu (余聪)} 
\affiliation{School of Physics and Astronomy, Sun Yat-Sen University, Zhuhai 519082, China}
\affiliation{CSST Science Center for the Guangdong-Hong Kong-Macau Greater Bay Area, Zhuhai 519082, China}

\author[0000-0001-9290-7846]{Ruobing Dong (董若冰)}
\affiliation{Kavli Institute for Astronomy and Astrophysics, Peking University, Beijing 100871, China}
\affiliation{Department of Physics and Astronomy, University of Victoria, Victoria, BC, V8P 5C2, Canada}

\correspondingauthor{Xiaoyi Ma, Ruobing Dong}
\email{xiaoyima@stu.pku.edu.cn, rbdong@pku.edu.cn}

\begin{abstract}

Observations of protoplanetary disks have revealed the presence of both crescent-shaped and ring-like structures in dust continuum emission. These crescents are thought to arise from dust-trapping vortices generated by the Rossby Wave Instability (RWI), which induces density waves akin to those caused by planets. These vortices have the potential to create gaps and rings within the disk, resulting from the dissipation of their density waves. We carry out 2D hydrodynamic simulations in the shearing box to investigate vortex-disk interaction. 
We find that long-lived vortices can produce dust rings and gaps in inviscid discs detectable by ALMA, and a more elongated vortex produces rings at larger separations. Vortex-induced density waves carry over two orders of magnitude higher angular momentum flux compared to planet-induced ones that shock at the same location, making the former much more effective at producing dust gaps and rings far away.

\end{abstract}

\section{Introduction}

Recent observations have revealed dust rings and gaps in protoplanetary disks (e.g., \citealt{ALMA_2015, Andrews_2018, Long_2018}). The accumulation of dust within these rings can accelerate grain growth, making them significant sites for planetesimal formation \citep{Sierra_2019}. Understanding their formation mechanisms is crucial for comprehending disk evolution and planetary formation.

Various mechanisms have been proposed to form observed rings in protoplanetary disks \citep{Bae_2023}. Among them, a popular explanation is the gravitational interaction between young planets and protoplanetary disks \citep[e.g.,][]{Dong_2015}. Density waves are excited by planets at Lindblad resonances, and they carry angular momentum away as they propagate. When waves dissipate, their carried angular momentum is deposited to local disk material, resulting in gap opening \citep{Lin_1986}. Meanwhile, gas pressure bumps form at the inner and outer gap edges, and dust drifts towards pressure bumps to form rings \citep{Pinilla_2012}. 

While dozens of planets in disks have been predicted based on observed rings and gaps \citep{Zhang_2018, Bae_2018b, Lodato_2019}, 
they have largely escaped detections due to the formidable planet-to-star flux ratios \citep{Asensio-Torres_2021} 
. Whether planets are responsible for the majority of dust rings and gaps remains debated \citep{Jiang_2021}.
In addition, dust rings and gaps have been found in disks so young that the local material has only had time to complete on the order of 1,000 orbits \citep{Segura-Cox_2020}. Whether planets can form and generate disk structures so quickly remains unclear.

In this study, we aim to explore a novel and non-planetary mechanism for generating dust rings and gaps in disks through vortex-disk interactions. Vortices are anticyclonic structures characterized by gas streamlines revolving around their center. They may form through the Rossby Wave Instability (RWI), initially identified via linear analysis \citep{Lovelace_1999, Li_2000} and further explored in its nonlinear evolution through simulations \citep{Li_2001, Meheut_2013a}. RWI is triggered by rapid radial variations in vortensity in low-turbulence discs. Such conditions may be realized at edges of gaps \citep{Zhu_2014,Miranda_2017} or narrow rings \citep{Lovelace_1999, Li_2000}. Vortices can trap dust particles sufficiently and concentrate solids at their center \citep{Meheut_2012c,Meheut_2013b,Meheut_2013c}, resulting in crescent-shaped structures observable in continuum emission \citep{Birnstiel_2013, Baruteau_2016}. Such structures have been found in many disks \citep{van_der_Marel_2013, van_der_Marel_2020, Casassus_2013}.

In disks, vortices can induce velocity perturbations, and excite spiral density waves \citep{Li_2001, Meheut_2013a}. 
While previous works have focused on the excitation of the RWI and the subsequent vortex formation, this study takes a step forward by investigating the substructures that emerge from vortex-disk interactions. Similar to the ones excited by planets, vortex-induced density waves may also propagate and dissipate, leading to gap opening. Candidate vortices and dust rings have been found to coexist in some disks, such as HD 135344B \citep{van_der_Marel_2016} and HD 143006 \citep{Perez_2018, Andrews_2018}, promoting the hypothesis that the two may be physically related.

We carry out hydrodynamics simulations to study the interactions between a non-migrating vortex and an inviscid disk. The paper is organized as follows. We introduce our simulations in \S\ref{Method} and present their results in \S\ref{Result}, before discussing our findings in \S\ref{Discussion}. We summarize the main finding in \S\ref{Conlusion}.

\section{Method}\label{Method}

We conduct 2D hydrodynamics simulations of protoplanetary disks in the shearing box using the Athena++ code \citep{Stone_2020} to examine the propagation and dissipation of vortex-generated density waves, as well as the subsequent gap-opening process. We employ the simulations for isothermal disks. Viscosity, self-gravity and magnetic field are not included, and their effects will be discussed in \S\ref{other_effects}. The numerical setup, the initial conditions, and the boundary conditions are specified in \S\ref{section:setup}, \S\ref{section:initial}, and \S\ref{section:BC}, respectively, with a list of models in Table~\ref{table:1}.

\subsection{Numerical Setup}\label{section:setup}

The shearing box approximation \citep{Narayan_1987, Stone_2010} adopts a frame of reference at the radius $r_0$ co-rotating with the disk at the orbital velocity $\Omega_0 = \Omega(r_0)$. In this frame, we define a Cartesian coordinate system ($x, y, z$) with the unit vector ($\mathbf {\hat{i}}$, $\mathbf{\hat{j}}$, $\mathbf {\hat{k}}$) as:
\begin{equation}\label{eqn:shearing_box}
\begin{aligned}
x &= r -r_0\\
y &= r_0(\phi - \Omega_0 t)\\
z &= z
\end{aligned}
\end{equation}
where ($r, \phi, z$) are cylindrical coordinates. The orbital motion in the shearing box is approximated by a linear function as $-q\Omega_0x\mathbf{\hat{j}}$, where the shearing parameter $q$ is defined as:

\begin{equation}\label{eqn:omega}
q \equiv  -\frac{d \log \Omega}{d \log r}
\end{equation}
For  Keplerian flow, $q=3/2$. The 2D hydrodynamical equations are written in the co-rotating Cartesian coordinate system as: 
\begin{equation}\label{eqn:NS-eqn_gas_1}
    \frac{\partial \Sigma_\text{gas}}{\partial t}+ \nabla \cdot [\Sigma_\text{gas}\mathbf{v_\text{gas}}] = 0,
\end{equation}
\begin{equation}
    \begin{aligned}
    \frac{\partial {\Sigma_\text{gas}\mathbf{v_\text{gas}}}}{\partial t}+ \nabla \cdot [\Sigma_\text{gas} \mathbf{v_\text{gas}}\mathbf{v_\text{gas}}+P] = \\
    \Sigma_\text{gas}\Omega_0^2(2qx\mathbf{\hat{i}})-2\Omega_0\mathbf{\hat{k}}\times\Sigma_\text{gas}\mathbf{v_\text{gas}}
    \end{aligned}\label{eqn:NS-eqn_gas_2}
\end{equation}
where \ref{eqn:NS-eqn_gas_2} are the gas continuity and momentum equation, respectively, where $\Sigma_\text{gas}$ is the gas surface density, $\mathbf{v_\text{gas}}$ is the gas velocity and $P$ is the gas pressure. For isothermal disks, $P = c_s^2\Sigma_\text{gas}$, where $c_s$ is isothermal sound speed.

We include a dust fluid module in the Athena++ \citep{Huang_2022} by treating dust as pressureless fluid. The continuity and momentum equations for dust are defined as follows:
\begin{equation}\label{eqn:NS-eqn_dust_1}
    \frac{\partial \Sigma_\text{dust}}{\partial t}+ \nabla \cdot [\Sigma_\text{dust} \mathbf{v_\text{dust}]} = 0,
\end{equation}
 \begin{equation}
    \begin{aligned}
    \frac{\partial {\Sigma_\text{dust}\mathbf{v_\text{dust}}}}{\partial t}+ \nabla \cdot [\Sigma_\text{dust} \mathbf{v_\text{dust}}\mathbf{v_\text{dust}}] = \\
    \Sigma_\text{dust}\Omega_0^2(2qx\mathbf{\hat{i}})-2\Omega_0\mathbf{\hat{k}}\times\Sigma_\text{dust}\mathbf{v_\text{dust}}+ \Sigma_\text{dust}\frac{\mathbf{v_\text{gas}}-\mathbf{v_\text{dust}}}{t_s}
    \end{aligned}\label{eqn:NS-eqn_dust_2}
\end{equation}
The symbols are defined similarly to the gas fluid. The stopping time, $t_s$, represents the timescale over which the velocity of dust particles exponentially decays due to aerodynamic drag. We only include one dust species with a Stokes number $St = 0.1$, neglecting dust feedback and diffusion. The Stokes number $St$  of dust particles in the Epstein regime is:

\begin{equation}\label{eqn:Stoke_number}
    St = \frac{\pi s \rho_\mathrm{m,dust}}{2\Sigma_\mathrm{gas}}
\end{equation}
where $s$ is the size of the dust particle and $\rho_\mathrm{m,dust}$ is the material density of the dust particle. While $St$, instead of $s$, is fixed in our simulations, the gas surface density $\Sigma_\mathrm{gas}$ varies only on the order of 10\%, resulting in a near-constant $s$.

We utilize the second-order piecewise linear method (PLM) spatial reconstruction and the Van-Leer 2 (VL2) time integrator, the HLLE Reimann solver for gas and the default Reimann solver for dust~\citep{Huang_2022} in this study. To reduce computational costs, we incorporate the orbital advection algorithm to speed up the azimuthal calculations and reduce the trancation errors. \citep{Masset_2000,Stone_2020}. 

The grid is uniformly spaced, and the cells are squared. Based on the convergence test in Appendix \ref{resolution}, a resolution of 256 cells per disk scale height ($h$) is required to accurately model weak density wave propagation and shocks in an inviscid environment. This is in line with previous results \citep{Dong_2011a}, as lower resolutions may lead to numerical diffusion and premature wave dissipation. \par

All simulations are carried out for a minimum of 400 orbits, corresponding to 0.4 Myr at a distance of 100 AU around a solar mass star.

\begin{table*}
\caption{List of Disk Models}
\label{table:1}
\resizebox{\linewidth}{!}{%
\renewcommand{\arraystretch}{1.2}
\hspace{-3.5cm}
\begin{tabular}{c|c c c| c c c c c c| c c}
\hline
 & \multicolumn{3}{c|}{Initial Setup$^{\rm a}$} & \multicolumn{6}{c|}{Model Characteristics$^{\rm b}$} &\multicolumn{2}{c}{Planet Analog$^{\rm c}$}\\
\hline\hline
Column & 1 & 2 & 3 & 4 & 5 & 6 & 7 & 8 & 9 & 10 & 11\\
\hline
& Bump      & \multicolumn{2}{c|}{Simulation Domain} & Most Unstable & Characteristic & Aspect & Wave      & AMF & Ring     &  \multicolumn{2}{c}{Planet mass}  \\
Name & Amplitude & \multicolumn{2}{c|}{}                  & Mode          & Time            & Ratio & Amplitude &     & Position & \multicolumn{2}{c}{} \\

 & $A_0$ & $(x_\text{min}, x_\text{max})$ & $(y_\text{min}, y_\text{max})$  & $m_*$ &$\tau_{20}$ & $\chi_\mathrm{v}$ & $(\delta\Sigma_\mathrm{gas})_\mathrm{max}$ & $F_J$ & $x_\text{ring}$ & $M_{\rm p,\mathrm{shock}}$ & $M_{\rm p,\mathrm{AMF}}$\\ 
\hline \hline
 Epoch & & & & & & $t = \tau_{20}$ & $t$ = 20 orbits &  $t$ = 20 orbits & $t = \tau_{20}$ & &\\
 \hline
 Unit & $[\Sigma_\mathrm{gas,0}]$ & [$h$] &[$h$] & &[orbits]& &$[\Sigma_\mathrm{gas,0}]$  & ($\times 10^{-5}$) & [$h$] &[$M_\mathrm{th}$] &[$M_\mathrm{th}$]\\
 \hline \hline
 A1 & 0.439 &$[-20, 20]$ & $[-10\pi/4, 10\pi/4]$ & 4 & 149 & 4.00 & 0.25 & 2.87 & 8.16 & 0.011 & 0.176\\ 
 A2 & 0.358 & $[-20, 20]$ & $[-10\pi/4, 10\pi/4]$  & 4 &  210 & 4.61 & 0.18 & 1.40 & 9.15 & 0.006 & 0.123\\ 
 A3 & 0.227 & $[-30, 30]$ & $[-10\pi/3, 10\pi/3]$  & 3 &  461 & 5.99 & 0.14 & 0.61 & 12.33 & 0.002 & 0.081\\ 
 A4 & 0.205  & $[-30, 30]$ & $[-10\pi/3, 10\pi/3]$ & 3 &  860 & 8.28 & 0.04 & 0.10 & 16.88 & 0.001 & 0.034\\
 \hline
\end{tabular}%
}

{\raggedright$^{\rm a}$: The initial setup of the models is described by the density contrast amplitude of the initial bump ($A_0$) and simulation domain ($(x_\text{min}, x_\text{max})$, $(y_\text{min}, y_\text{max})$).\par $^{\rm b}$:The parameters that characterized the models are the following (see \S\ref{section:relation}):\par 
\begin{itemize}
\item$m_*$: the most unstable mode of RWI calculated by \cite{Ono_2016, Ono_2018}
\item$\tau_{20}$: characteristic time when $\langle\delta\Sigma_{\rm dust}/\Sigma_{\rm dust,0}\rangle_y\sim$ 20 at the location of dust ring B 
\item $\chi_\mathrm{v}$: aspect ratio of the vortex measured at $t = \tau_{20}$
\item$( \delta\Sigma_\mathrm{gas})_\mathrm{max}/\Sigma_\mathrm{gas,0}$: peak wave amplitude at its shock location when it is initially excited at $t$ = 20 orbits
\item$F_J$: angular momentum flux carried by the wave at $t$ = 20 orbits
\item$x_\mathrm{ring}$: location of the ring further away from the vortex at $t = \tau_{20}$ 
\end{itemize}
$^{\rm c}$:$M_{\rm p,\mathrm{shock}}$ and $M_{\rm p,\mathrm{AMF}}$ are the masses of the planet that excites waves which shock at the same location and carry the same amount of AMF as the vortex (see \S\ref{section:planet}).}\par
\end{table*}

\subsection{Initial conditions}\label{section:initial}
Our main goal is to study vortex-disk interaction rather than the vortex-formation process. While the RWI can be triggered in a variety of scenarios, such as at the dead zone edge \citep{Miranda_2017} or at the edge of planetary gaps \citep{Zhu_2014}, we choose to trigger RWI by introducing a Gaussian surface density bump to the uniform background density, following \cite{Ono_2018}. This method generates vortices akin to those produced by other mechanisms. We adopt this approach as it allows us to isolate the vortex-disk interaction from other complications, such as viscosity variation near the dead zone and density waves induced by planets.

The initial gas surface density profile is given by:
\begin{equation}\label{eqn:background_density}
    \Sigma_{\text{gas}} (x,y) = \Sigma_{\text{gas},0} \left(1+A_0e^{-\frac{1}{2}\left(\frac{x-x_0}{\Delta w_0}\right)^2}\right)
\end{equation}
where $\Sigma_{\text{gas},0}=1$ is the uniform background profile. The initial dust surface density follows the gas with a uniform dust-to-gas ratio of 0.01.
The gas background profile is characterized by three parameters: $A_0$ for the density contrast of the bump, $\Delta w_0$ for its radial half-width, and $x_0$ for the location of the bump center. 

We fix $x_0 = 0$ and $w_0 = 0.632h$. Here, $h = c_s/\Omega_0 = 0.1$, with $\Omega_0 = 1$ and isothermal sound speed $c_s = 0.1$ in code units. We explore the parameter space for $A_0$ (column 1 in Table~\ref{table:1})  using four disk models. Each model corresponds to a most unstable azimuthal mode, $m_*$ (column 4 in Table~\ref{table:1}). These values are determined through linear analysis of RWI \citep{Ono_2016, Ono_2018}.

The initial velocities are set by hydrostatic equilibrium:
\begin{eqnarray}\label{eqn:initial_velocity}
v_\text{gas,$x$,0} &=& 0\\
v_\text{gas,$y$,0} &=& -q\Omega_0x+\frac{1}{2\Omega_0\Sigma_{\text{gas}}}\frac{\mathrm{d} P}{\mathrm{d}x}
\end{eqnarray}

In addition, we impose a small initial gas velocity perturbation in $x$ according to the most unstable azimuthal mode $m_*$ to trigger the RWI: 

\begin{equation}\label{eqn:velocity_per}
\delta v_\text{gas,$x$,0} = 10^{-5}\cos{(m_*y)} e^{-\frac{1}{2}(\frac{x}{0.1})^2}
\end{equation}

\subsection{Domain size and boundary conditions (BC)} \label{section:BC}

The simulation domain in the $x$ direction ($x_\text{min},x_\text{max}$) (column 2 in Table~\ref{table:1}) is set to fully capture the shock dissipation for all disk models, which will be elaborated in \S\ref{section:relation}.\par

The simulation domain in the $y$ direction ($y_\text{min},y_\text{max}$) (column 3 Table~\ref{table:1}) is determined by its most unstable azimuthal mode $m_*$. According to \cite{Ono_2018}, the number of primary vortices formed equals $m_*$, which later merges into a single vortex. To avoid complications associated with vortex merging, we truncate the full azimuthal domain to $1/m_*$, leaving only one vortex within the domain throughout the simulation. 

The boundary condition (BC) is periodic in the $x$ direction and shearing periodic in the $y$ direction. To minimize the wave reflection along $x$ edges, we add a wave-damping zone along the $x$ boundaries to gradually damp all variables $X$ (gas and dust density, and velocities) to the initial values by the following equation \citep{Huang_2022}:
\begin{equation}\label{eqn:damping_zone}
\frac{\partial X}{\partial t} = \zeta_\text{dr}\Omega_0[X (t=0) - X]\left(\frac{|x-x_{\text{damping}}|}{\Delta x_\text{damping}}\right)^2,   
\end{equation}
where $\zeta_\text{dr}$ is the damping rate, which we set to be 100, and $x_{\text{damping}}$ denotes the damping edge, defined as $x_{\text{damping}}= 0.89x_{\text{max}}$ for the outer boundary and  $x_{\text{damping}}= 0.89x_{\text{min}}$ for the inner boundary. $\Delta x_\text{damping}$ represents the width of the damping zone, which is $\Delta x_\text{damping} = 0.11|x_{\text{max}}|$ for the outer boundary and $\Delta x_\text{damping}= 0.11|x_{\text{min}}|$ for the inner boundary.

\section{Result}\label{Result}

We start from the model with the lowest initial bump amplitude ({\fontfamily{lmtt}\selectfont Model A4}), showing gap opening in \S\ref{section:surface_density} and wave dissipation in \S\ref{section:wave}. We then move on to other models in Table \ref{table:1} and examine how vortex-disk interactions depend on vortex properties in \S\ref{section:relation}. 
Finally, we compare the locations of the gaps opened by vortices and planets, as well as the angular momentum flux carried by their density waves in \S\ref{section:planet}.

\subsection{Density wave excitation and gap opening by the vortex} \label{section:surface_density}

\begin{figure*}
\begin{center}
\includegraphics[width=\textwidth]{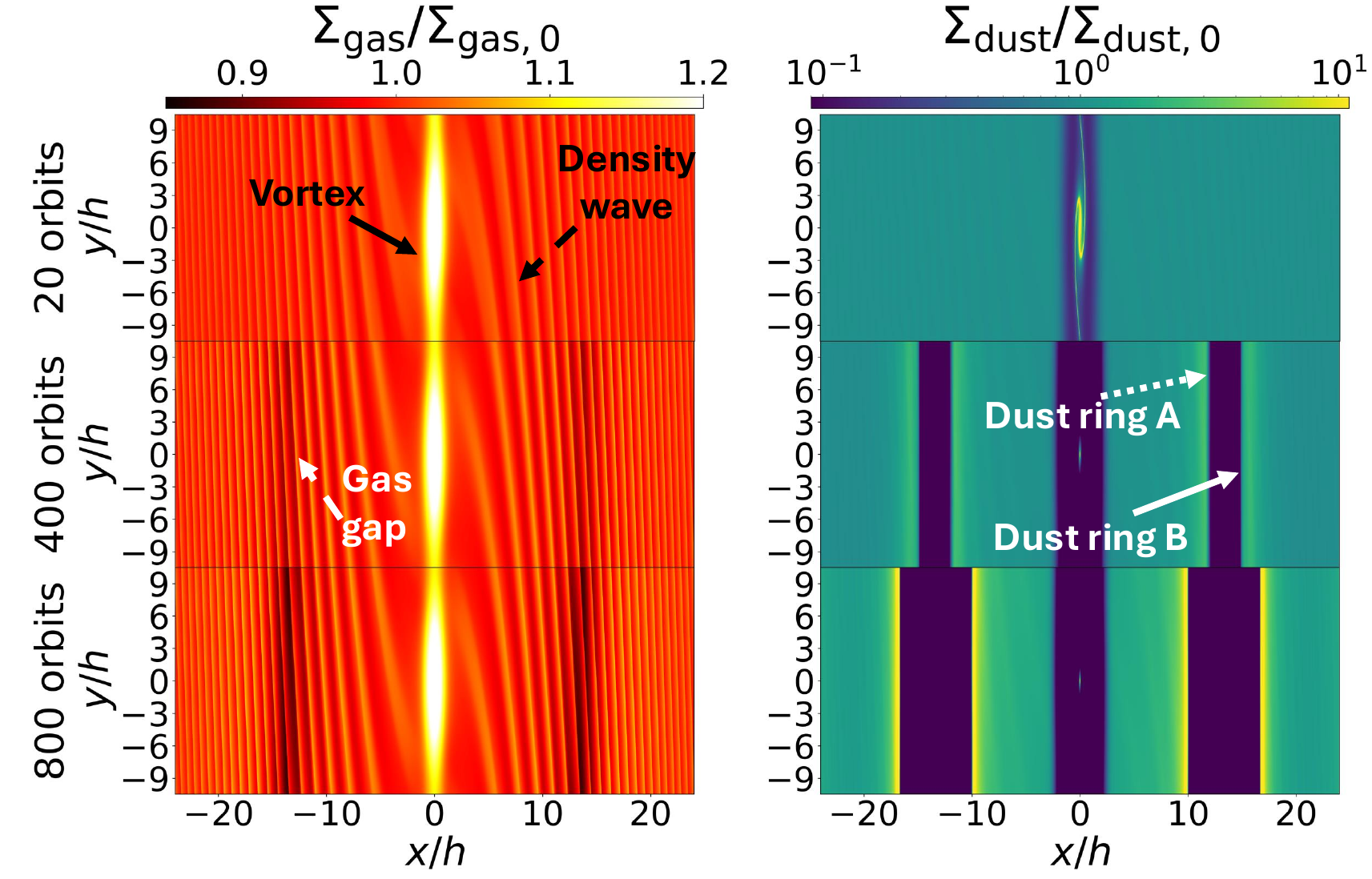}
\caption{Surface density snapshots for gas (left) and dust (right) of module with lowest initial bump amplitude ({\fontfamily{lmtt}\selectfont Model A4}), at $t$ = 20, 400 and 800 orbits. An RWI vortex centred at $(x,y)=(0,0)$ forms (black solid arrow, left column), generating spiral density waves (black dashed arrow, left column). As the system evolves, a gas gap is opened on each side of the vortex (white dashed arrow, left column), prompting the formation of two dust rings at its two edges (dust ring A and B; white dotted arrow and white solid arrow, right column). The gas gap deepens and widens with time, pushing the dust rings further apart. See \S \ref{section:surface_density} for details.}
\label{fig:A4_surface_density}
\end{center}
\end{figure*}

Figure \ref{fig:A4_surface_density} presents snapshots of the gas (left) and dust surface density (right) for {\fontfamily{lmtt}\selectfont Model A4} in Table \ref{table:1}, illustrating the evolution of the system. At the start of the simulation, the gas density bump undergoes RWI. A vortex (black solid arrow, left column) then forms within tens of orbits. One vortex excites a pair of density waves (black dashed arrow, left column) on each side\footnote{The two waves on each side are symmetric if the vortex is symmetric (\cite{Huang_2019} Figure 1). In our simulations with periodic BC in $y$, since the vortex is as long as the simulation domain, the two waves overlap.}, typically with low contrasts ($\sim10\%$ in this case), as shown in \cite{Huang_2019}. A gas gap (white dashed arrow, left column) gradually forms on each side as the waves dissipate, prompting the formation of two pressure bumps at its two edges. \par
Due to the aerodynamic gas drag, dust particles drift towards the pressure bumps and vortex. A dust gap is opened at the location of the gas gap, and two dust rings (white solid arrow and white dotted arrow, right column) form at its two pressure bumps. The gas gap deepens and widens with time, pushing the two dust rings further apart.

The central position of the gas (and dust) gap remains relatively unchanged. As there is one gap on each side of the vortex, in total there are two gaps and four dust rings in our simulation domain. The two dust rings closer to the vortex (dust ring A; white dotted arrow, right column) move towards each other and may eventually merge into the vortex, as observed in the models with higher initial density contrast $A_0$ (see \S\ref{section:relation}). We will only focus on the dust ring further away from the vortex (dust ring B; white solid arrow, right column).

\subsection{Shock Development and Wave Dissipation}\label{section:wave}

We investigate the origin of the gaps in {\fontfamily{lmtt}\selectfont Model A4} by studying the propagation and dissipation of the density waves excited by the vortex.

\begin{figure}
\includegraphics[width=0.48\textwidth]{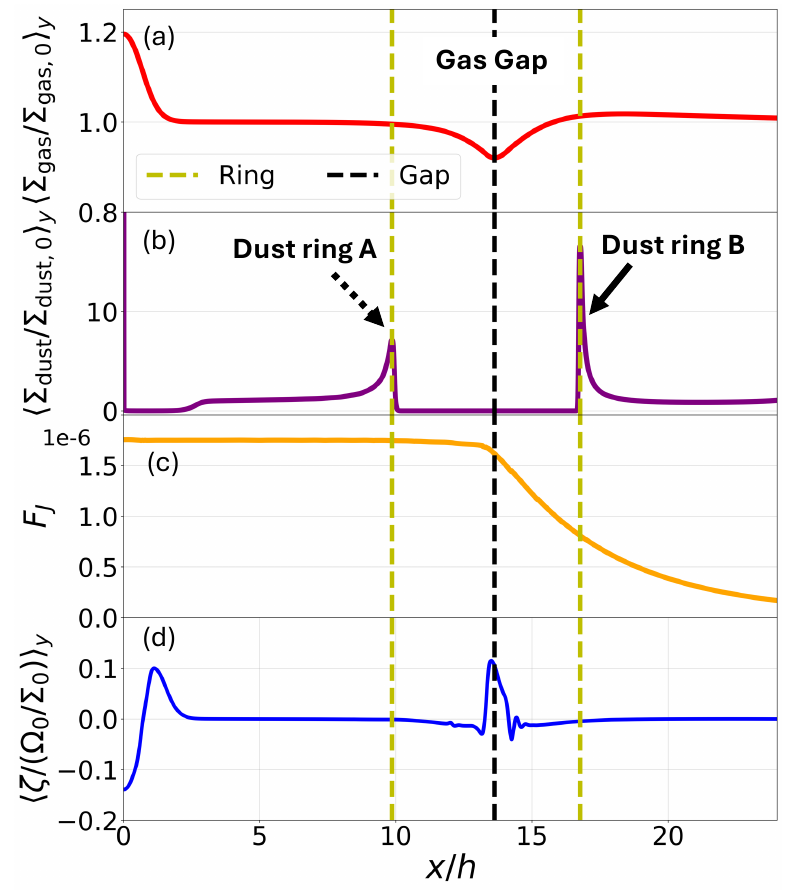}
\caption{The azimuthally averaged radial profiles for the gas (a) and dust surface density (b), the angular momentum flux carried by the waves (c), and vortensity (d) for {\fontfamily{lmtt}\selectfont Model A4} at $t = 800$ orbits (bottom panels in Figure~\ref{fig:A4_surface_density}). Since the shearing box is symmetric with respect to the $y$ axis, this image only displays the profiles for $x>0$. The peaks at $x/h$ = 0 in panels (a) and (b) are over-densities at the center of the vortex. The black dashed line at $x=13.6h$ marks the location of the gas gap trough (a), the starting point of the AMF decay (c), and the vortensity peak (d), indicating the shock in the wave. The two yellow dashed lines indicate the locations of the two dust rings in panel (b). We will focus on dust ring B that is located at $16.9h$ away from the vortex (solid arrow).}
\label{fig:AMF_vortensity}
\end{figure}

We calculate the angular momentum flux (AMF) carried by the density wave \citep{Miranda_2020}:
\begin{equation}\label{eqn:AMF}
  F_{J}(x) = \int_{y_\text{min}}^{y_\text{max}} \Sigma_\text{gas}(x,y)v_{\text{gas},x}(x,y)\delta v_{\text{gas},y}(x,y) dy,
\end{equation}
where $v_{\text{gas},x}(x,y)$ is the radial velocity and $\delta v_{\text{gas},y}(x,y)=v_{\text{gas},y}(x,y)-v_{\text{gas},y,0}(x,y)$ is the azimuthal velocity perturbation.
The total AMF acquired by the wave during its excitation is determined by vortex-disk interactions and is a characteristic of the vortex.
Before the wave shocks, its AMF remains constant. Once it shocks, the angular momentum carried by the wave is gradually transferred to the local disk, causing the AMF to decay with distance $x$ \citep{Goodman_2001}, as illustrated in Figure \ref{fig:AMF_vortensity}(c). Where the AMF starts to decay thus indicates the shock location.
As this transfer of angular momentum causes gap opening, the AMF acquired by the wave prior to dissipation sets its total gap opening ability.

The shock location can also be identified by searching for the jump in \textit{potential vorticity}, also known as \textit{vortensity} \citep{Cimerman_2021}:
\begin{equation}\label{eqn:vortensity}
\zeta \equiv \frac{(\nabla \times \mathbf{v_\text{gas}})_z}{\Sigma_\text{gas}}.
\end{equation}
Vortensity remains conserved in inviscid and barotropic environment, except at shock locations, where vortensity is excited \citep{Kevlahan_1997, Li_2005}.
 
Figure \ref{fig:AMF_vortensity} displays the azimuthally averaged radial profiles of normalized gas surface density (panel (a)), dust surface density (panel (b)), AMF carried by the density waves (panel (c)), and vortensity (panel (d)) for {\fontfamily{lmtt}\selectfont Model A4} at 800 orbits. Since the shearing box is symmetric with respect to the $y$ axis, we will focus on discussing one side of the simulation. 

The gas gap is situated at a distance of 13.6$h$ from the vortex, denoted by the black dashed line. There are two narrow dust rings (yellow dash lines, panel (b) ) at the two pressure bumps at the gap edges. Dust ring A (dotted arrow, panel (b)) located $9.9h$ away from the vortex and dust ring B (solid arrow, panel (b)) is $16.9h$ away.
The gas gap location (black dashed line, panel (a)) aligns with the jump in vortensity (black dashed line, panel (d)), and the starting point of the AMF decay (black dashed line, panel (c)), indicating that the gas gap is indeed opened by wave dissipation caused by the shock. We thus conclude that a vortex can create rings and gaps through its density waves.

\begin{figure}
\begin{center}
\includegraphics[width=0.48\textwidth]{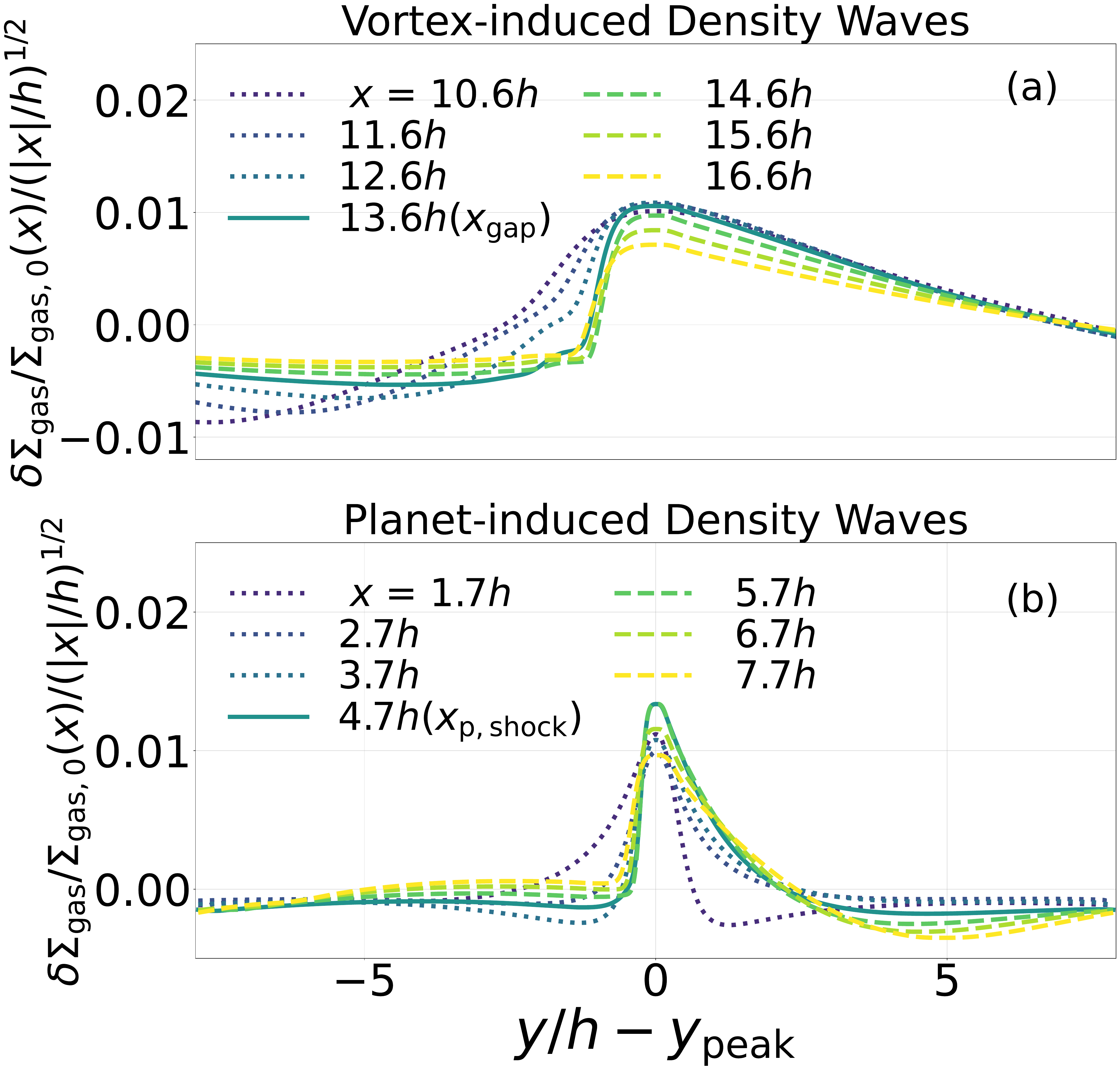}
\caption{Panel (a): Azimuthal profiles for vortex-induced density wave of {\fontfamily{lmtt}\selectfont Model A4} at $t$ = 20 orbits (top left panel in Figure~\ref{fig:A4_surface_density}). The shock location is identified in Figure~\ref{fig:AMF_vortensity} (black dashed line) as $x_\mathrm{gap} = 13.6h$. The profiles are depicted at seven radii ($x$), including three pre-shock locations ($10.6h$, $11.6h$ and $12.6h$) (dotted lines), one shock location (solid line), and three post-shock locations ($14.6h$, $15.6h$ and $16.6h$) (dashed lines). (b): Azimuthal profiles for planet-induced density waves ($M_{\rm p} = 0.01 M_{\rm th}$) at $t$ = 20 orbits. The shock location $x_{\rm p, \rm shock} = 4.7h$ is calculated using Equation \ref{eqn:planet_mass_shock}. The profiles are depicted at seven radii ($x$), including three pre-shock locations ($1.7h$, $2.7h$ and $3.7h$) (dotted lines), one shock location (solid line), and three post-shock locations ($5.7h$, $6.7h$ and $7.7h$) (dashed lines).}
\label{fig:A4_wave_profle}
\end{center}
\end{figure}

Next, we examine the azimuthal wave profiles, $\delta\Sigma_\text{gas} = \Sigma_\text{gas} - \Sigma_{\text{gas},0}$ as a function of $y$ at a few different $x$, for {\fontfamily{lmtt}\selectfont Model A4} in Figure \ref{fig:A4_wave_profle}.  The profiles are aligned by their peaks $y_\text{peak}$. We choose an early frame at $t$ = 20 orbits to minimize the influence of gap opening. 
As the density wave travels radially away from the vortex, its amplitude increases with distance $x$ as $x^{1/2}$ due to the conservation of AMF \citep{Goodman_2001}.
To compensate for this, we normalize the wave profiles by $x^{1/2}$. 
The leading edge of the wave gradually steepens, and becomes nearly vertical at $x=13.6h$, where the wave shocks. Before that, the $x^{1/2}-$normalized wave amplitude is conserved; after that, the amplitude drops due to wave dissipation. 
The excitation--shock--dissipation process of the waves generated by a vortex resembles that of the waves generated by a planet \citep{Muto_2010,Dong_2011b}. However, the two types of waves have different profiles, with the planet induced ones having a full-width-half-magnitude $\sim$70\% smaller, and slightly increase in wave amplitude when the wave shocks (Figure \ref{fig:A4_wave_profle}(b)). In Section \ref{section:planet}, we compare the properties of waves excited by vortices with those excited by planets.

\subsection{Comparative Analysis of Vortex Properties and Substructures in Disk Models}\label{section:relation}

We analyze vortex and ring properties in different models (Table~\ref{table:1}) to identify the correlation between the vortex aspect ratio and the location of the dust ring. From now on, ``dust ring" specifically refers to dust ring B (white solid arrow, Figure \ref{fig:A4_surface_density}), the one further from the vortex, unless otherwise specified. 

\begin{figure*}
\centering
\includegraphics[width=\textwidth]{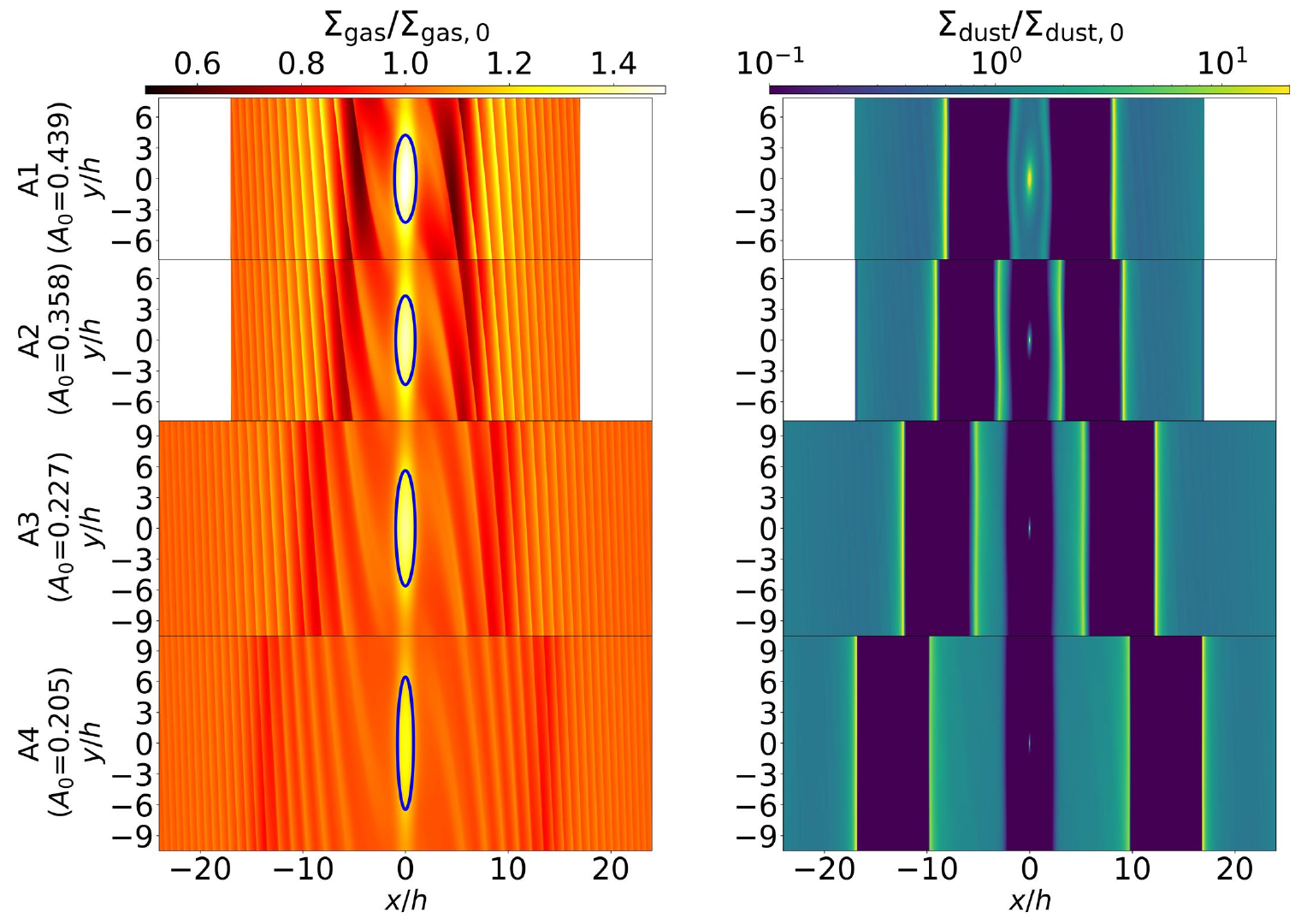}
\caption{Surface density for gas (left) and dust (right) at $t = \tau_{20}$ for disk models listed in Table \ref{table:1}. The blue ellipse delineates the edge of the vortex core. Models with a lower initial density contrast $A_0$ exhibit more elongated vortices, excite density waves with lower amplitudes that shock at larger distances, and produce gaps and rings situated further away from the vortex.}
\label{fig:disk_models}
\end{figure*}

\begin{figure}[h]
\includegraphics[width=0.48\textwidth]{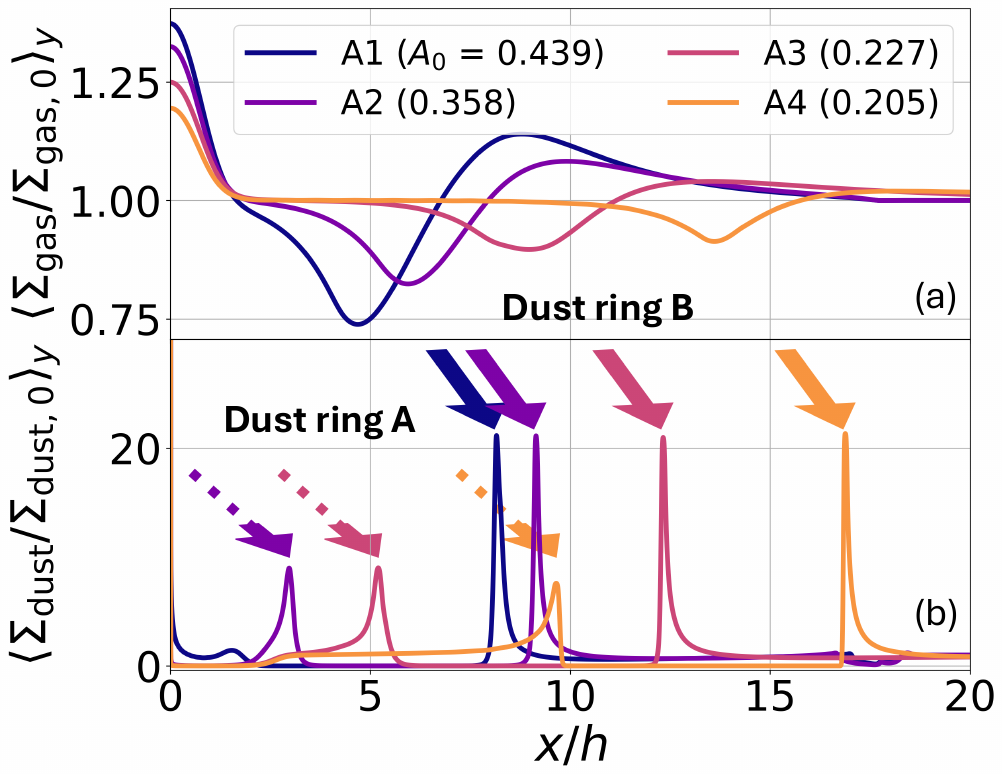}
\caption{Azimuthally averaged surface density profiles for gas (top; left column in Figure~\ref{fig:disk_models}) and dust (bottom; right column in Figure~\ref{fig:disk_models}) at $t = \tau_{20}$ (column 5 in Table~\ref{table:1}). Models with a higher $A_0$ feature a gas gap closer to the vortex (top), resulting in closer dust rings as well (bottom). The dust ring closer to the vortex in each pair (dust ring A; the dotted arrows in each dotted$+$solid pair) eventually moves into the vortex, as seen in {\fontfamily{lmtt}\selectfont Model A1}.}

\label{fig:model_profile}
\end{figure}

\begin{figure}
\includegraphics[width=0.48\textwidth]{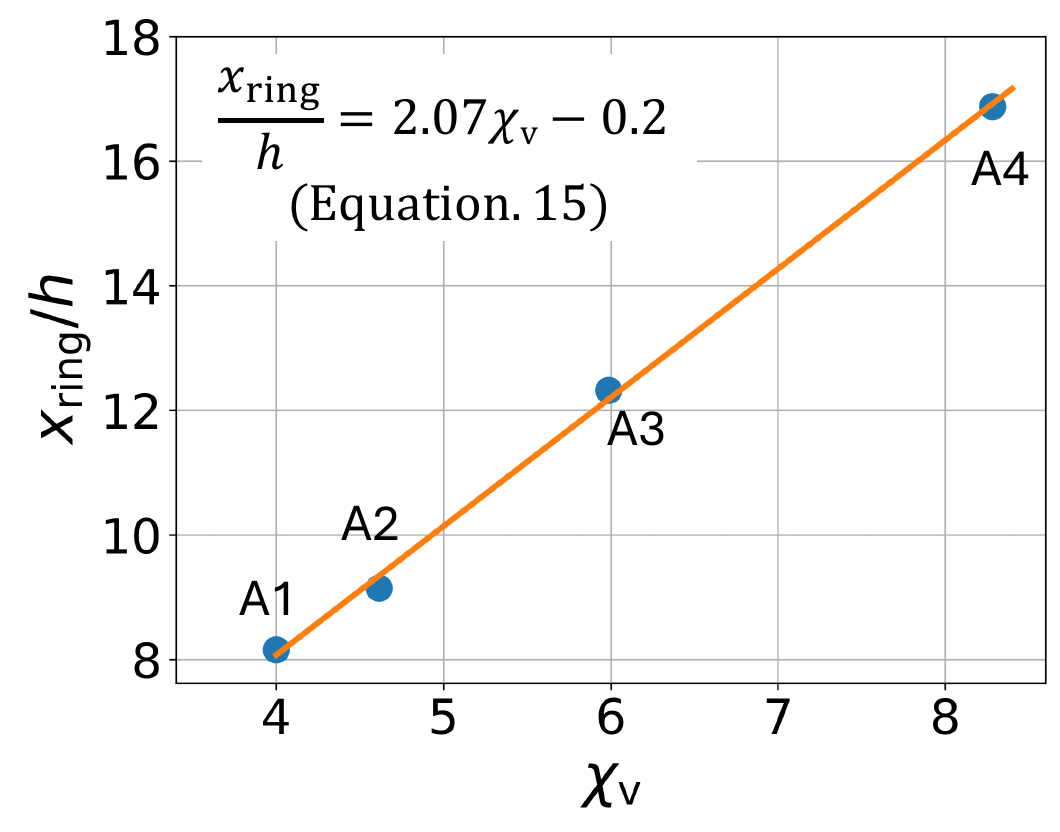}
\caption{Empirical fitting between the aspect ratio of the vortex and the location of the dust ring.}
\label{fig:curve_fit}
\end{figure}

Models with a lower initial surface density contrast $A_0$ produce more elongated vortices (blue ellipse, left column in Figure \ref{fig:disk_models}), quantified by the aspect ratio $\chi_\mathrm{v}$ (column 6 in Table \ref{table:1}) as defined in Appendix \ref{aspect_ratio}. Such vortices excite density waves with lower amplitudes ($(\delta\Sigma_\mathrm{gas})_\mathrm{max}$; column 7 in Table \ref{table:1}) and that carry less AMF ($F_J$; column 8 in Table \ref{table:1}). These waves shock at larger distances \citep{Goodman_2001}, producing gaps and rings ($x_{\mathrm{ring}}$; column 9 in Table \ref{table:1}) further away from the vortex (Figure~\ref{fig:disk_models}). This trend is further illustrated in Figure~\ref{fig:model_profile}, showing the locations of the gaps and rings in azimuthally averaged surface density profiles. More quantitatively, Figure~\ref{fig:curve_fit} shows that $\chi_\mathrm{v}$ is in a tight linear correlation ($\chi^2 = 0.007$) with $x_\mathrm{ring}$, the location of the dust ring in the pair produced by each gas gap (solid arrows in Figure \ref{fig:model_profile}, bottom):
\begin{equation}\label{eqn:empircal_relation}
    x_\mathrm{ring}/h = a\chi_\mathrm{v} - b,
\end{equation}
where $a = 2.07 \pm 0.06$ and $b = -0.2\pm 0.3$.

\subsection{Comparison with Planet-disk Interaction}\label{section:planet}

Both planets and vortices can excite density waves that propagate in disks, which shock and dissipate at certain distances to open gaps. However, the profiles of the two categories of waves are not the same. This is evident by comparing panel (a) with panel (b) in Figure~\ref{fig:A4_wave_profle}, which shows the evolution of weak waves induced by a sub-thermal mass planet. As the propagation of planet-induced waves and the associated gap opening process have been thoroughly studied \citep{Goodman_2001, Dong_2011b, Dong_2011a}, we hope to compare the two to help us understand vortex-induced waves. Specifically, we determine the mass of the corresponding planets whose waves shock at the same distance, or carry the same AMF, as those induced by the vortex.

\begin{figure*}
\includegraphics[width=0.95\textwidth]{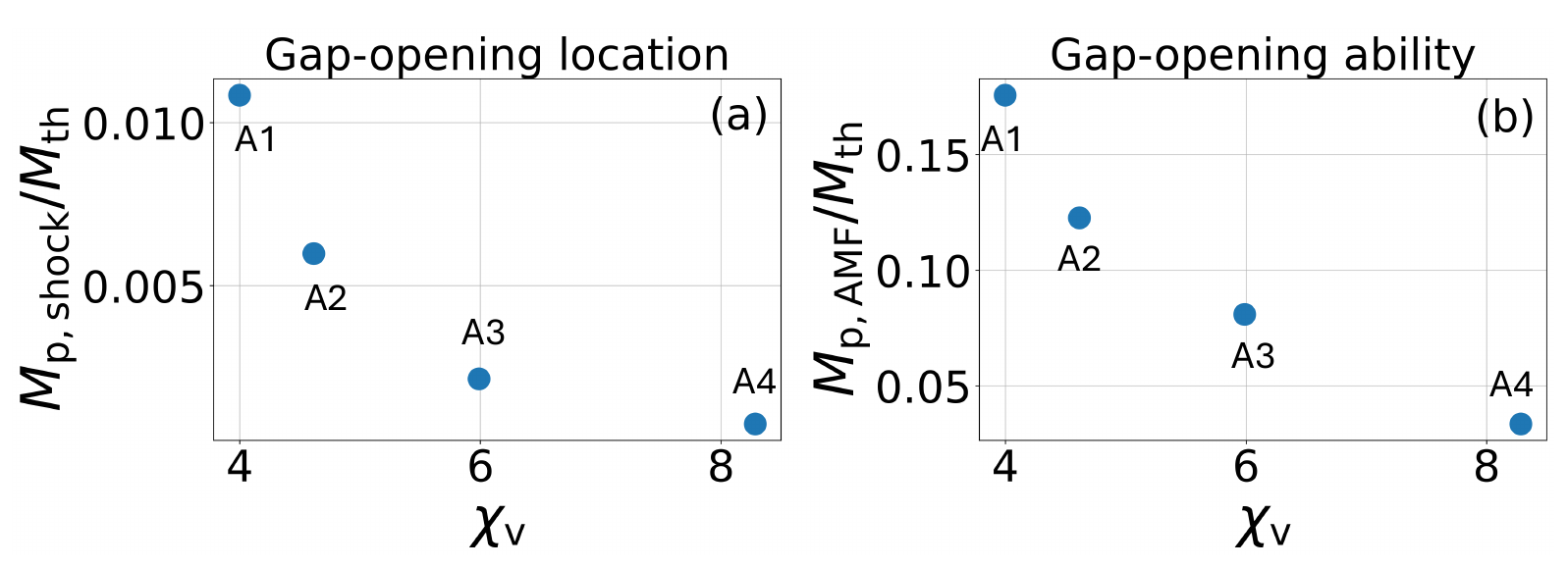}
\caption{Panel (a): Correlation between vortex aspect ratio and planet mass for the planets whose waves shock at the same distance as the vortex-induced ones (See \S \ref{section:planet}). (b): Correlation between vortex aspect ratio and planet mass for the planets whose waves carry the same amount of AMF as the vortex-induced ones (See \S \ref{section:planet}).}
\label{fig:planet}
\end{figure*}

In Figure~\ref{fig:planet}(a) we show the mass of the planet whose primary density wave\footnote{A low mass planet can excite multiple density waves that shock at different locations \citep{Dong_2017,Bae_2018a}. Here we refer to the primary wave with the shortest shock length.} shock at the same distance as vortex-induced wave for each of our models using the relation found by \citet{Goodman_2001}:
\begin{equation} \label{eqn:planet_mass_shock}
   M_{\rm p, \rm shock}\approx\left(\frac{|x|_{\rm shock}}{0.8h}\right)^{-5/2}\left(\frac{12/5} {\gamma+1}\right)M_\mathrm{th}
\end{equation}
where $|x|_{\rm shock}$ is the shock length of the vortex-induced wave, corresponds to the location of the gas gap (See \S \ref{section:wave} for details), $\gamma = 5/3$ is the adiabatic index and $M_\mathrm{th} = c_s^3/G\Omega_{\rm p}$ is the disk thermal mass \citep{Goodman_2001, Dong_2011b}. For a disk with $h/r$ = 0.1 around a solar mass star at the planetary location, the thermal mass will be roughly 1 Jupiter mass \citep{Miranda_2019}. Vortices in all our models correspond to sub-thermal mass planets ($M_{\rm p, \rm shock} \ll M_{\rm th}$), while less elongated vortices correspond to more massive planets.

Since waves excited by a planet and a vortex differ in profile, those with the same shock length may carry different AMF. In Figure~\ref{fig:planet}(b) we find the mass of the planet whose density waves carry the same AMF as vortex-induced ones ($F_{J}$; column 8 in Table \ref{table:1}) for each of our models  using the relation in Equation \ref{eqn:planet_AMF} \citep{Dong_2011b}:
\begin{equation}\label{eqn:planet_AMF}
    M_{\rm p, \rm AMF}\approx \left(\frac{F_{J}\Omega_{\rm p}}{0.93\Sigma_0r_{\rm p}{c_s}^3}\right)^{1/2}M_\mathrm{th}.
\end{equation}
Observationally, the waves excited by such a planet have the same gap opening potential as the ones excited by the corresponding vortex. Less elongated vortices that generate density waves with higher AMF correspond to more massive planets. 

\begin{figure}
\includegraphics[width=0.48\textwidth]{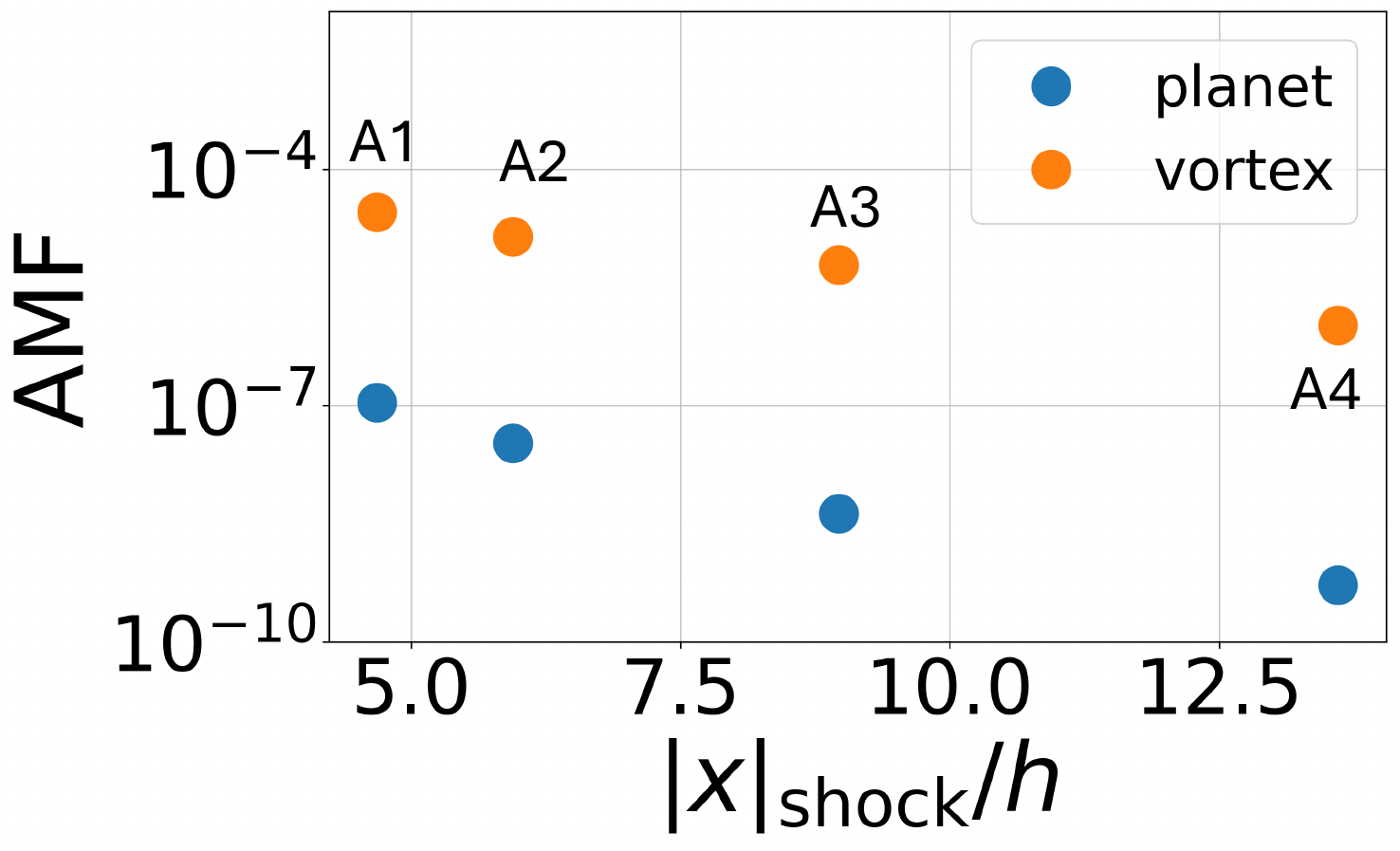}
\caption{AMF carried by the planet-induced waves (blue dots) and vortex-induced waves (orange dots) that shock at the same distance.}
\label{fig:planet_AMF}
\end{figure}

Combining two panels in Figure~\ref{fig:planet}, we see that $M_{\rm p, \rm AMF}$ can be one order of magnitude higher than $M_{\rm p, \rm shock}$. For a planet whose waves shock at the same distance as those induced by a vortex, 
the AMF carried by the vortex-induced waves (orange dots, Figure~\ref{fig:planet_AMF}) can be two orders of magnitude higher than the AMF carried by the planet-induced waves (blue dots,  Figure~\ref{fig:planet_AMF}), indicating the greater gap-opening ability of the vortex compared with the planet.

\section{Discussion} \label{Discussion}
In this section, we compare the synthetic image of vortex-disk interaction with real observations (\S\ref{observation}) and discuss the impact of physical effects not considered in our study (\S\ref{other_effects}).

\subsection{Comparison with Real Observation}\label{observation}

\begin{figure*}
\includegraphics[width=\textwidth]{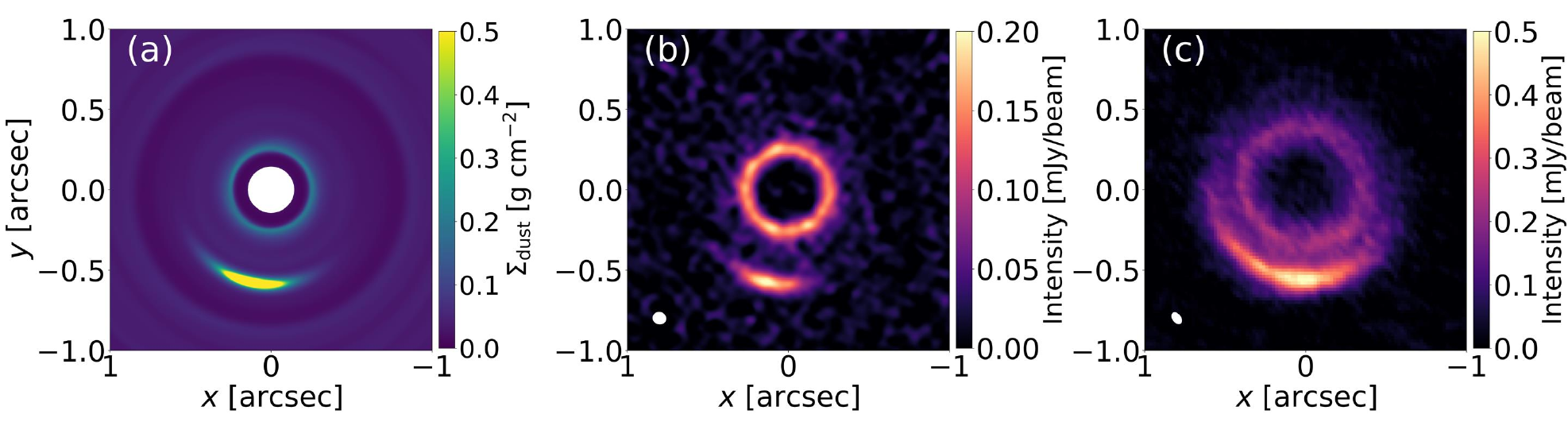}
\caption{Panel (a): Dust surface density from the global simulation for the disk {\fontfamily{lmtt}\selectfont Model A2}. (b): The corresponding synthetic continuum image at 1.9 mm. The white ellipse in the bottom left corner represents the synthetic beam size of 0.079$''$ ×0.070$''$. 
(c): Continuum observations for HD135344B at 1.9 mm \citep{Cazzoletti_2018}. The white ellipse in the bottom left corner indicates the beam size of 0.076$''$ ×0.047$''$.}
\label{fig:synthetic_image}
\end{figure*}

 While recent observations have revealed that several disks exhibit azimuthal asymmetries \citep{van_der_Marel_2020}, only a few disks hold both asymmetries and dust rings, such as HD 135344B \citep[Figure~\ref{fig:synthetic_image}(c);][]{Cazzoletti_2018}.\par

HD 135344B features an inner dust cavity, an inner ring and an outer azimuthal asymmetry in dust continuum. Our synthetic image (Figure \ref{fig:synthetic_image}(b)), created as described in Appendix \ref{Image}, resembles these observed features, suggesting the vortex may have produced the inner dust ring.

\subsection{Other Physical Effects}\label{other_effects}

Our simulations exclude physical effects such as vortex migration, viscosity, radiative cooling, self-gravity, 3D dynamics, and dust feedback. Below, we discuss how these factors may influence the contrast of vortex-induced substructures and the vortex lifetime.

\subsubsection{Contrast of Vortex-Induced Substructures}\label{Viscosity}

While our simulations assume an inviscid disk, viscous processes can influence the gap-opening process by smoothing disk structures, thereby counteracting gap formation \citep{Fung_2014}. Thus, the density contrast of the rings and gaps in a viscous disk would be shallower compared with those within Figure \ref{fig:A4_surface_density} and Figure \ref{fig:disk_models}. 

Figure \ref{fig:synthetic_image} shows a synthetic image from a simulation for a locally isothermal disk. In disks with finite cooling times, the density contrast of rings and gaps varies with the cooling timescales of the disks. This could cause the ring in Figure \ref{fig:synthetic_image} to appear either brighter or dimmer compared to those in disks with finite cooling times \citep{Zhang_2023}. This is because the angular momentum flux is not conserved in locally isothermal disks, as the waves exchange angular momentum with the local disk as they propagate \citep{Miranda_2019}.

\subsubsection{Vortex Longevity} \label{Vortex_Longevity}

In a global disk, vortices generate asymmetric density waves on the two sides due to geometric effects, causing radial migration of the vortex similar to the Type I migration of planets \citep{Paardekooper_2010}. The migration timescale is usually comparable to or longer than the disk lifetime unless the aspect ratio drops below 4 or the vortex exceeds a radial size of $2h$.

Relatively low viscosity ($\alpha \lesssim 10^{-4}$) is required to sustain a long-lived vortex \citep{Godon_1999, Fu_2014b}. The high viscosity in hot disks can dampen the perturbations with small azimuthal mode numbers \citep{Gholipour_2014}. Observations from ALMA infer low $\alpha$ values for the disk, with upper limits ranging from $10^{-4}$ to $10^{-3}$, suggesting that the maintenance of long-lived vortices is feasible in real disks \citep{Pinte_2023}. 

Radiative cooling can also impact the lifetime of the vortex. \cite{Fung_2021} found that the decay of the vortex is slow in both the isothermal and adiabatic limits, where the gas cooling time is much shorter or much longer than the vortex turnaround time.

RWI vortices exhibit vertical velocities in both 3D numerical simulations \citep{Meheut_2010} and linear calculations \citep{Meheut_2012a}, with 3D growth rates comparable to those in 2D. Long-term simulations indicate that some RWI vortices can persist and grow over hundreds of years if the disk sustains the conditions that trigger RWI, such as dead zones \citep{Meheut_2012b}. Nevertheless, elliptical instability can disrupt vortices with small aspect ratios ($\chi \leq 4$) \citep{Lesur_2009, Richard_2013} in 3D, whereas more elongated vortices behave comparably in both 3D and 2D scenarios. 

Self-gravity inhibits long-lived vortex formation \citep{Pierens_2018}, but it may aid 3D vortices in resisting elliptical instabilities and extend their lifetime \citep{Lin_2018}. 

Dust-to-gas feedback might also impact vortex lifetime. When the dust-to-gas mass ratio surpasses 30$\%$ to 50$\%$ within the vortex, the dust feedback may excite dynamical instability and destroy the vortex \citep{Fu_2014b, Crnkovic_2015}. 

Overall, various effects can either limit or extend vortex lifetimes. In more realistic scenarios involving multiple effects, vortex evolution and its corresponding disk interaction remain open areas for future research.

\section{Summary}\label{Conlusion}

We found that vortices can create dust rings and gaps in inviscid protoplanetary disks through shock dissipation of their induced density waves (Figure \ref{fig:A4_surface_density}). By conducting 2D gas+dust simulations with varying vortex shapes, we established a linear correlation $x_\mathrm{ring}/h \sim 2.07 \chi_\mathrm{v}$ between vortex aspect ratio ($\chi_\mathrm{v}$) and dust ring location ($x_\mathrm{ring}$) (Equation \ref{eqn:empircal_relation}; Figures~\ref{fig:disk_models} and \ref{fig:curve_fit}). More elongated vortices generate density waves that carry less angular momentum flux (AMF), producing rings at larger separations. 

In comparison to planet-disk interactions, we found that vortex-induced density waves carry over two orders of magnitude more AMF than planet-induced waves with the same shock length, making vortices significantly more effective at gap opening (Figure \ref{fig:planet}).

Our study introduces a novel, non-planetary mechanism for the formation of rings and gaps in protoplanetary disks via vortex-disk interactions. Further research in more realistic disk environments is crucial to refine our understanding of this new mechanism.

\section*{Acknowledgements}
We are grateful to an anonymous referee for constructive suggestions that improved our paper. We would like to thank Doug Johnstone, Jeffrey Fung, Josh Calcino, Gianluigi Bodo, Gregory Herczeg, Haochang Jiang, Kengo Tomida, Min-Kai Lin, Nienke van der Marel, Roman Rafikov, Rixin Li and Tomohiro Ono for their comments and suggestions. We appreciate the Chinese Center for Advanced Science and Technology for hosting the Protoplanetary Disk and Planet Formation Summer School in 2022, organized by Xue-ning Bai and Ruobing Dong, which facilitated this research. X.M., P.H., and R.D. acknowledge the support of the Natural Sciences and Engineering Research Council of Canada (NSERC) [RGPIN-2023-05299], and the Alfred P. Sloan Foundation. P. H. acknowledges the financial support from NSERC ALLRP 577027-22. This research was enabled in part by support provided by the Digital Research Alliance of Canada alliance.can.ca.

\appendix
\section{Convergence Test in Vortensity}\label{resolution}

To confirm if a long-lived vortex can produce substructures through shock dissipation of density waves, it is crucial for our simulations to have sufficient resolution to capture the shock location, which is identified by peaks in vortensity (\S\ref{section:wave}). Achieving convergence in vortensity is therefore essential for our goal.

\begin{figure}
\centering
\includegraphics[width=0.8\textwidth]{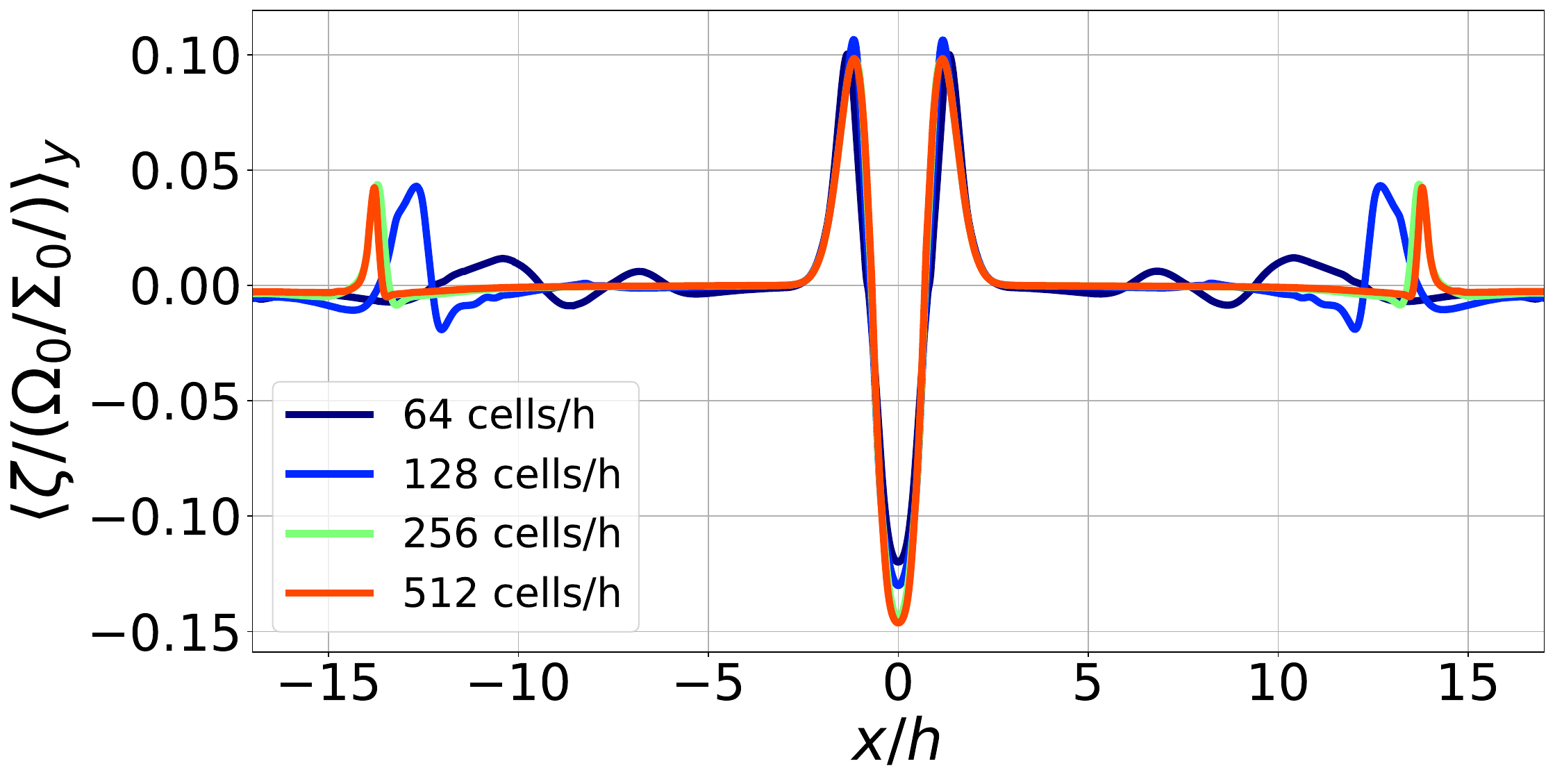}
\caption{Vortensity radial profile at $t = 360$ orbits for {\fontfamily{lmtt}\selectfont Model A4} is shown at resolutions of 64, 128, 256 and 512 cells/h.}
\label{fig:A.1}
\end{figure}

Figure \ref{fig:A.1} illustrates the radial profiles of vortensity for {\fontfamily{lmtt}\selectfont Model A4} at various resolutions: 64, 128, 256, and 512 cells/h. We find that the location of the vortensity peaks converged at 256 cells/h. Thus, 256 cells/h is adequate for our simulations and analysis.

\section{Vortex Shape}\label{aspect_ratio}

\begin{figure}[h]
\centering
\includegraphics[width=0.8\textwidth]{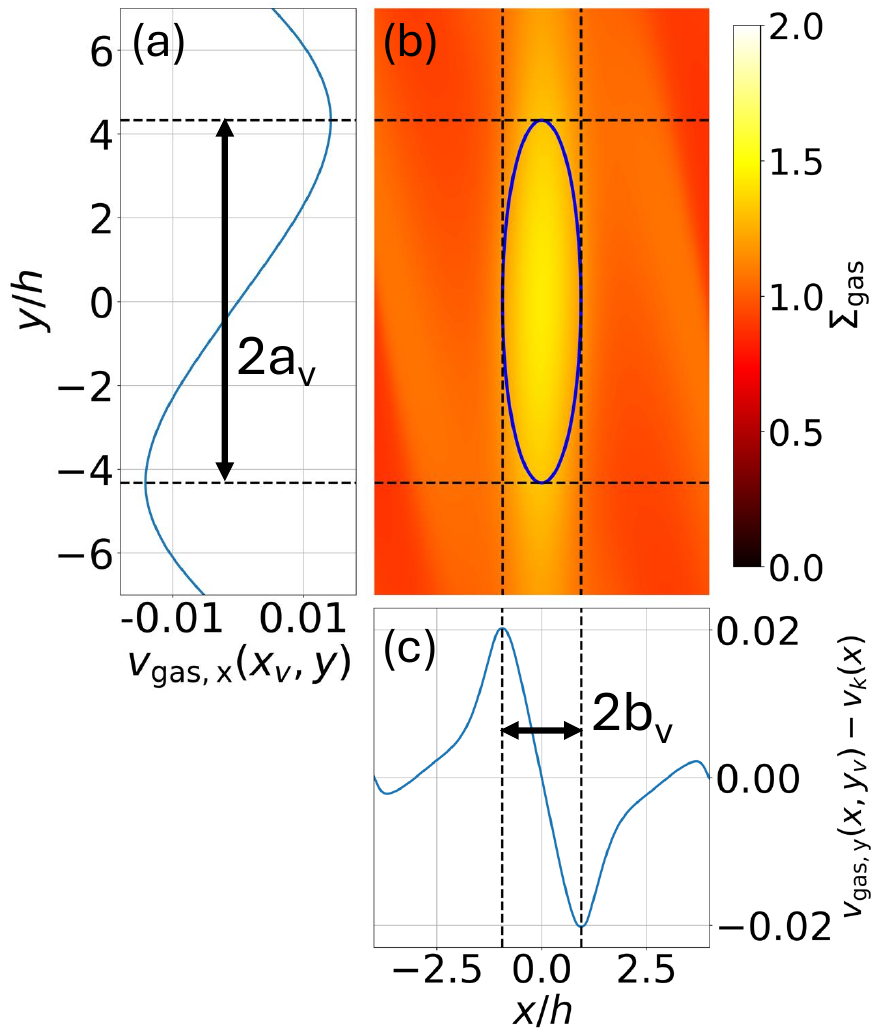}
\caption{Radial velocity $v_{\rm{gas},x}$ along the $y$ direction (a), surface gas density (b), and azimuthal velocity deviation from Keplerian velocity $v_{\text{gas},y}- v_k$ along the $x$ direction (c) for {\fontfamily{lmtt}\selectfont Model A2} at $t = \tau_{20}$. Black dashed lines mark local extrema near the vortex. The distance between local maximum and minimum in $v_{\rm gas, x}$ defines the semi-major axis $a_\text{v}$, while for $v_{\text{gas},y}- v_k$ it defines the semi-minor axis $b_\text{v}$. An ellipse, delineated by $a_\text{v}$ and $b_\text{v}$ from the velocity profile, represents the vortex shape in blue.}
\label{fig:aspect_ratio}
\end{figure}

The shape of the vortex is characterized by its aspect ratio, $\chi_\text{v} = a_\text{v}/b_\text{v}$. The semi-major ($b_\text{v}$) and semi-minor axis ($a_\text{v}$) of the vortex core are calculated from velocity profiles ($v_{\text{gas},y}(x,y_\text{v}) - v_k(x)$, $v_{\text{gas},x}(x_\text{v},y)$) near the vortex ($x_v$,$y_v$), where $v_k(x)$ is Keplerian velocity.

Figure \ref{fig:aspect_ratio} shows these velocity profiles for {\fontfamily{lmtt}\selectfont Model A4} at $t = \tau_{20}$. Both profiles converge to zero at the vortex center and exhibit variations with distance from the vortex center, attaining their local extremes near the edge of the vortex core.  Hence, $a_\text{v}$ and $b_\text{v}$ is estimated as half the distance between two extremas in $v_{\text{gas},y}(x,y_\text{v}) - v_k(x)$ and $v_{\text{gas},x}(x_\text{v},y)$, respectively. The vortex in Figure \ref{fig:aspect_ratio} exhibits an aspect ratio of $\chi_\text{v} = 4.61$.

\section{Synthetic observations}\label{Image}

\subsection{Simulation}
We use the grid-based hydrodynamics code Athena++ to conduct 2D global simulations in polar coordinates ($r$,$\phi$) for generating synthetic images. We consider the equation of state of the gas to be locally isothermal, where the radial profile of temperature is $T(r) \propto (r/r_0)^{-0.5}$, where $r_0 = 65$ AU. The corresponding radial profile for the scale height of the gas disk $h$ is $h/r = 0.1(r/r_0)^{0.25}$. Here, $h = c_s/\Omega$, where $c_s$ is the sound speed. We introduce viscosity with $\alpha = 10^{-5}$ \citep{Shakura_1973}, and incorporate dust diffusion into the simulations. 

To trigger RWI, we impose the pressure bump defined in Equation \ref{eqn:background_density} at 1.5 $r_0$ using the parameters of {\fontfamily{lmtt}\selectfont Model A2}. The background density follows a power law, $\Sigma_{\text{gas}}(t = 0) = \Sigma_{\text{gas,0}}(r/r_0)^{-0.5}$, where $\Sigma_{\text{gas,0}}= 5.7$ g $\mathrm{cm^{-2}}$. The total disk mass is 0.06 $M_{\odot}$.

Our simulation domain spans 0.1 $r_0$ to 3.2 $r_0$ radially with constant logarithmic spacing and covers 0 to $2\pi$ uniformly in the azimuthal direction. We employ a root mesh grid of $62 \times 192$, with level 4 refinement between 0.25 $r_0$ and 2.5 $r_0$ near the vortex to capture density wave evolution. Each level of refinement doubles the resolution, so the resolution in the radial direction at $r_0$ reaches 33 cells per gas scale height at the finest grids.

The initial velocities are set by hydrostatic equilibrium for gas. Small velocity perturbations in white noise with an amplitude of $0.01c_s$ are added in both radial and azimuthal directions to induce instability.

We model one dust species with $St = 0.01$, with an initial uniform dust-to-gas ratio of 0.01. At $t = 0$ and $r = r_0$, $s = 0.3$ mm for $\rho_\mathrm{m,dust}$ = 1.2 g cm$^{-3}$. To prevent rapid inward drift at the start of the simulation, dust motion begins at $t = 200$ orbits.

Periodic boundary conditions are enforced along the $\phi$ direction, while the radial boundary condition is kept fixed to the initial condition, following the damping zone described by Equation \ref{eqn:damping_zone} with a damping rate of $\zeta_\text{dr} = 1$. The damping edge is calculated as $1.3^{-2/3}x_{\text{max}}$ for the outer boundary and $2.0^{2/3}x_{\text{min}}$ for the inner boundary.

\subsection{Radiative Transfer and Synthetic Observations}\label{Global:4.2}

We follow the procedure outlined in \citep{Huang_2020} to post-process the outputs of 2D hydrodynamic global simulation, using the radiation transfer code RADMC-3D \footnote{\url{https://www.ita.uni-heidelberg.de/~dullemond/software/radmc-3d/}}\citep{Dullemond_2012} and the Common Astronomy Software Application (CASA) \footnote{\url{https://casa.nrao.edu/}}\citep{The_CASA_Team_2022}. The simulation is initialized using the stellar parameters of the disk HD 135344B \citep{van_der_Marel_2016}, assuming the central star is a luminous blackbody sphere with $R_{\star}$ = 2.2 $R_{\odot}$, $M_{\star}$ = 1.6 $M_{\odot}$ and $T$ = 6590 K.

As RADMC-3D reads 3D density distributions instead of surface densities, we convert the output 2D surface densities to 3D density distributions using the following profile, assuming hydrostatic equilibrium for a vertically isothermal disk \citep{Armitage_2020}:
\begin{equation}\label{eqn:3D_density}
  \rho (r,\phi, z) = \frac{1}{\sqrt{2\pi}}\frac{\Sigma(r,\phi)}{h(r)}e^{-z^2/2h(r)^2}
\end{equation}
Here we consider two kinds of dust. Submicron-sized dust particles scatter and absorb stellar light at the disk surface. These dust particles are well-coupled with gas with a constant dust-to-gas of 0.01 and the same scale height as gas. 

We obtain the disk temperature using \texttt{mctherm} from RADMC-3D, assuming the temperature is primarily influenced by sub-micron-sized dust particles. The dust opacity input for RADMC-3D is computed using optool\footnote{\url{https://github.com/cdominik/optool}}  \citep{Dominik_2021}, employing the DSHARP composition \citep{Birnstiel_2018} and a power-law dust size distribution given by $n(a) \propto a^{-3.5}$, where $a$ represents the grain size. The minimum grain size considered is 0.01 $\mu$m, while the maximum grain size is set to 100 $\mu$m. Additionally, we assume that the dust grains are solid spheres without any porosity.

The dust emission at a wavelength of 1.9 mm is determined by utilizing 0.3 mm dust from the hydrodynamic simulation through ray-tracing in RADMC-3D. The dust surface density from the hydrodynamic simulation is converted to 3D dust density distribution using Equation \ref{eqn:3D_density}, assuming the dust scale height to be 10$\%$ of the gas scale height. The opacity of 0.3 mm dust is also computed by employing the DSHARP composition \citep{Birnstiel_2018}. Additionally, we include the effect of dust scattering in our computations.

To simulate the ALMA observation, we process the noise-free image from RADMC-3D using \texttt{simobserve} in CASA to incorporate observational noise and convolve it with the appropriate beam size. We adopt the ALMA configuration of C4.8, similar to the real observation of HD 135344B \citep{Cazzoletti_2018}, resulting in a synthetic beam size of $0.079" \times 0.070"$. The synthetic image is generated with an integration time of 2 hours.

\bibliographystyle{aasjournal}
\bibliography{references}

\begin{thebibliography}{}
\expandafter\ifx\csname natexlab\endcsname\relax\def\natexlab#1{#1}\fi
\providecommand{\url}[1]{\href{#1}{#1}}
\providecommand{\dodoi}[1]{doi:~\href{http://doi.org/#1}{\nolinkurl{#1}}}
\providecommand{\doeprint}[1]{\href{http://ascl.net/#1}{\nolinkurl{http://ascl.net/#1}}}
\providecommand{\doarXiv}[1]{\href{https://arxiv.org/abs/#1}{\nolinkurl{https://arxiv.org/abs/#1}}}

\bibitem[{{ALMA Partnership} {et~al.}(2015){ALMA Partnership}, {Brogan}, {P{\'e}rez}, {Hunter}, {Dent}, {Hales}, {Hills}, {Corder}, {Fomalont}, {Vlahakis}, {Asaki}, {Barkats}, {Hirota}, {Hodge}, {Impellizzeri}, {Kneissl}, {Liuzzo}, {Lucas}, {Marcelino}, {Matsushita}, {Nakanishi}, {Phillips}, {Richards}, {Toledo}, {Aladro}, {Broguiere}, {Cortes}, {Cortes}, {Espada}, {Galarza}, {Garcia-Appadoo}, {Guzman-Ramirez}, {Humphreys}, {Jung}, {Kameno}, {Laing}, {Leon}, {Marconi}, {Mignano}, {Nikolic}, {Nyman}, {Radiszcz}, {Remijan}, {Rod{\'o}n}, {Sawada}, {Takahashi}, {Tilanus}, {Vila Vilaro}, {Watson}, {Wiklind}, {Akiyama}, {Chapillon}, {de Gregorio-Monsalvo}, {Di Francesco}, {Gueth}, {Kawamura}, {Lee}, {Nguyen Luong}, {Mangum}, {Pietu}, {Sanhueza}, {Saigo}, {Takakuwa}, {Ubach}, {van Kempen}, {Wootten}, {Castro-Carrizo}, {Francke}, {Gallardo}, {Garcia}, {Gonzalez}, {Hill}, {Kaminski}, {Kurono}, {Liu}, {Lopez}, {Morales}, {Plarre}, {Schieven}, {Testi}, {Videla}, {Villard}, {Andreani}, {Hibbard}, \&
  {Tatematsu}}]{ALMA_2015}
{ALMA Partnership}, {Brogan}, C.~L., {P{\'e}rez}, L.~M., {et~al.} 2015, \apjl, 808, L3, \dodoi{10.1088/2041-8205/808/1/L3}

\bibitem[{{Andrews} {et~al.}(2018){Andrews}, {Huang}, {P{\'e}rez}, {Isella}, {Dullemond}, {Kurtovic}, {Guzm{\'a}n}, {Carpenter}, {Wilner}, {Zhang}, {Zhu}, {Birnstiel}, {Bai}, {Benisty}, {Hughes}, {{\"O}berg}, \& {Ricci}}]{Andrews_2018}
{Andrews}, S.~M., {Huang}, J., {P{\'e}rez}, L.~M., {et~al.} 2018, \apjl, 869, L41, \dodoi{10.3847/2041-8213/aaf741}

\bibitem[{Armitage(2020)}]{Armitage_2020}
Armitage, P.~J. 2020, Astrophysics of Planet Formation, 2nd edn. (Cambridge University Press)

\bibitem[{{Asensio-Torres} {et~al.}(2021){Asensio-Torres}, {Henning}, {Cantalloube}, {Pinilla}, {Mesa}, {Garufi}, {Jorquera}, {Gratton}, {Chauvin}, {Szul{\'a}gyi}, {van Boekel}, {Dong}, {Marleau}, {Benisty}, {Villenave}, {Bergez-Casalou}, {Desgrange}, {Janson}, {Keppler}, {Langlois}, {M{\'e}nard}, {Rickman}, {Stolker}, {Feldt}, {Fusco}, {Gluck}, {Pavlov}, \& {Ramos}}]{Asensio-Torres_2021}
{Asensio-Torres}, R., {Henning}, T., {Cantalloube}, F., {et~al.} 2021, \aap, 652, A101, \dodoi{10.1051/0004-6361/202140325}

\bibitem[{{Bae} {et~al.}(2023){Bae}, {Isella}, {Zhu}, {Martin}, {Okuzumi}, \& {Suriano}}]{Bae_2023}
{Bae}, J., {Isella}, A., {Zhu}, Z., {et~al.} 2023, in Astronomical Society of the Pacific Conference Series, Vol. 534, Protostars and Planets VII, ed. S.~{Inutsuka}, Y.~{Aikawa}, T.~{Muto}, K.~{Tomida}, \& M.~{Tamura}, 423, \dodoi{10.48550/arXiv.2210.13314}

\bibitem[{{Bae} {et~al.}(2018){Bae}, {Pinilla}, \& {Birnstiel}}]{Bae_2018b}
{Bae}, J., {Pinilla}, P., \& {Birnstiel}, T. 2018, \apjl, 864, L26, \dodoi{10.3847/2041-8213/aadd51}

\bibitem[{{Bae} \& {Zhu}(2018)}]{Bae_2018a}
{Bae}, J., \& {Zhu}, Z. 2018, \apj, 859, 118, \dodoi{10.3847/1538-4357/aabf8c}

\bibitem[{Baruteau \& Zhu(2016)}]{Baruteau_2016}
Baruteau, C., \& Zhu, Z. 2016, Monthly Notices of the Royal Astronomical Society, 458, 3927–3941, \dodoi{10.1093/mnras/stv2527}

\bibitem[{{Birnstiel} {et~al.}(2013){Birnstiel}, {Dullemond}, \& {Pinilla}}]{Birnstiel_2013}
{Birnstiel}, T., {Dullemond}, C.~P., \& {Pinilla}, P. 2013, \aap, 550, L8, \dodoi{10.1051/0004-6361/201220847}

\bibitem[{Birnstiel {et~al.}(2018)Birnstiel, Dullemond, Zhu, Andrews, Bai, Wilner, Carpenter, Huang, Isella, Benisty, Pérez, \& Zhang}]{Birnstiel_2018}
Birnstiel, T., Dullemond, C.~P., Zhu, Z., {et~al.} 2018, The Astrophysical Journal Letters, 869, L45, \dodoi{10.3847/2041-8213/aaf743}

\bibitem[{{Casassus} {et~al.}(2013){Casassus}, {van der Plas}, {Perez}, {Dent}, {Fomalont}, {Hagelberg}, {Hales}, {Jord{\'a}n}, {Mawet}, {M{\'e}nard}, {Wootten}, {Wilner}, {Hughes}, {Schreiber}, {Girard}, {Ercolano}, {Canovas}, {Rom{\'a}n}, \& {Salinas}}]{Casassus_2013}
{Casassus}, S., {van der Plas}, G.~M., {Perez}, S., {et~al.} 2013, \nat, 493, 191, \dodoi{10.1038/nature11769}

\bibitem[{Cazzoletti {et~al.}(2018)Cazzoletti, van Dishoeck, Pinilla, Tazzari, Facchini, van~der Marel, Benisty, Garufi, \& Pérez}]{Cazzoletti_2018}
Cazzoletti, P., van Dishoeck, E.~F., Pinilla, P., {et~al.} 2018, Astronomy and amp; Astrophysics, 619, A161, \dodoi{10.1051/0004-6361/201834006}

\bibitem[{Cimerman \& Rafikov(2021)}]{Cimerman_2021}
Cimerman, N.~P., \& Rafikov, R.~R. 2021, Monthly Notices of the Royal Astronomical Society, 508, 2329–2349, \dodoi{10.1093/mnras/stab2652}

\bibitem[{{Crnkovic-Rubsamen} {et~al.}(2015){Crnkovic-Rubsamen}, {Zhu}, \& {Stone}}]{Crnkovic_2015}
{Crnkovic-Rubsamen}, I., {Zhu}, Z., \& {Stone}, J.~M. 2015, \mnras, 450, 4285, \dodoi{10.1093/mnras/stv828}

\bibitem[{{Dominik} {et~al.}(2021){Dominik}, {Min}, \& {Tazaki}}]{Dominik_2021}
{Dominik}, C., {Min}, M., \& {Tazaki}, R. 2021, OpTool: Command-line driven tool for creating complex dust opacities, Astrophysics Source Code Library, record ascl:2104.010

\bibitem[{{Dong} {et~al.}(2017){Dong}, {Li}, {Chiang}, \& {Li}}]{Dong_2017}
{Dong}, R., {Li}, S., {Chiang}, E., \& {Li}, H. 2017, \apj, 843, 127, \dodoi{10.3847/1538-4357/aa72f2}

\bibitem[{{Dong} {et~al.}(2011{\natexlab{a}}){Dong}, {Rafikov}, \& {Stone}}]{Dong_2011b}
{Dong}, R., {Rafikov}, R.~R., \& {Stone}, J.~M. 2011{\natexlab{a}}, \apj, 741, 57, \dodoi{10.1088/0004-637X/741/1/57}

\bibitem[{{Dong} {et~al.}(2011{\natexlab{b}}){Dong}, {Rafikov}, {Stone}, \& {Petrovich}}]{Dong_2011a}
{Dong}, R., {Rafikov}, R.~R., {Stone}, J.~M., \& {Petrovich}, C. 2011{\natexlab{b}}, \apj, 741, 56, \dodoi{10.1088/0004-637X/741/1/56}

\bibitem[{Dong {et~al.}(2015)Dong, Zhu, \& Whitney}]{Dong_2015}
Dong, R., Zhu, Z., \& Whitney, B. 2015, The Astrophysical Journal, 809, 93, \dodoi{10.1088/0004-637x/809/1/93}

\bibitem[{{Dullemond} {et~al.}(2012){Dullemond}, {Juhasz}, {Pohl}, {Sereshti}, {Shetty}, {Peters}, {Commercon}, \& {Flock}}]{Dullemond_2012}
{Dullemond}, C.~P., {Juhasz}, A., {Pohl}, A., {et~al.} 2012, RADMC-3D: A multi-purpose radiative transfer tool, Astrophysics Source Code Library, record ascl:1202.015

\bibitem[{{Fu} {et~al.}(2014){Fu}, {Li}, {Lubow}, \& {Li}}]{Fu_2014b}
{Fu}, W., {Li}, H., {Lubow}, S., \& {Li}, S. 2014, \apjl, 788, L41, \dodoi{10.1088/2041-8205/788/2/L41}

\bibitem[{Fung \& Ono(2021)}]{Fung_2021}
Fung, J., \& Ono, T. 2021, The Astrophysical Journal, 922, 13, \dodoi{10.3847/1538-4357/ac1d4e}

\bibitem[{Fung {et~al.}(2014)Fung, Shi, \& Chiang}]{Fung_2014}
Fung, J., Shi, J.-M., \& Chiang, E. 2014, The Astrophysical Journal, 782, 88, \dodoi{10.1088/0004-637x/782/2/88}

\bibitem[{Gholipour \& Nejad-Asghar(2014)}]{Gholipour_2014}
Gholipour, M., \& Nejad-Asghar, M. 2014, Monthly Notices of the Royal Astronomical Society, 441, 1910–1915, \dodoi{10.1093/mnras/stu697}

\bibitem[{{Godon} \& {Livio}(1999)}]{Godon_1999}
{Godon}, P., \& {Livio}, M. 1999, \apj, 523, 350, \dodoi{10.1086/307720}

\bibitem[{{Goodman} \& {Rafikov}(2001)}]{Goodman_2001}
{Goodman}, J., \& {Rafikov}, R.~R. 2001, \apj, 552, 793, \dodoi{10.1086/320572}

\bibitem[{Huang \& Bai(2022)}]{Huang_2022}
Huang, P., \& Bai, X.-N. 2022, The Astrophysical Journal Supplement Series, 262, 11, \dodoi{10.3847/1538-4365/ac76cb}

\bibitem[{Huang {et~al.}(2019)Huang, Dong, Li, Li, \& Ji}]{Huang_2019}
Huang, P., Dong, R., Li, H., Li, S., \& Ji, J. 2019, The Astrophysical Journal Letters, 883, L39, \dodoi{10.3847/2041-8213/ab40c4}

\bibitem[{Huang {et~al.}(2020)Huang, Li, Isella, Miranda, Li, \& Ji}]{Huang_2020}
Huang, P., Li, H., Isella, A., {et~al.} 2020, The Astrophysical Journal, 893, 89, \dodoi{10.3847/1538-4357/ab8199}

\bibitem[{Jiang \& Ormel(2021)}]{Jiang_2021}
Jiang, H., \& Ormel, C.~W. 2021, Monthly Notices of the Royal Astronomical Society, 505, 1162–1179, \dodoi{10.1093/mnras/stab1278}

\bibitem[{{Kevlahan}(1997)}]{Kevlahan_1997}
{Kevlahan}, N.~K.~R. 1997, Journal of Fluid Mechanics, 341, 371, \dodoi{10.1017/S0022112097005752}

\bibitem[{Lesur \& Papaloizou(2009)}]{Lesur_2009}
Lesur, G., \& Papaloizou, J. C.~B. 2009, Astronomy and amp; Astrophysics, 498, 1–12, \dodoi{10.1051/0004-6361/200811577}

\bibitem[{{Li} {et~al.}(2001){Li}, {Colgate}, {Wendroff}, \& {Liska}}]{Li_2001}
{Li}, H., {Colgate}, S.~A., {Wendroff}, B., \& {Liska}, R. 2001, \apj, 551, 874, \dodoi{10.1086/320241}

\bibitem[{{Li} {et~al.}(2000){Li}, {Finn}, {Lovelace}, \& {Colgate}}]{Li_2000}
{Li}, H., {Finn}, J.~M., {Lovelace}, R.~V.~E., \& {Colgate}, S.~A. 2000, \apj, 533, 1023, \dodoi{10.1086/308693}

\bibitem[{{Li} {et~al.}(2005){Li}, {Li}, {Koller}, {Wendroff}, {Liska}, {Orban}, {Liang}, \& {Lin}}]{Li_2005}
{Li}, H., {Li}, S., {Koller}, J., {et~al.} 2005, \apj, 624, 1003, \dodoi{10.1086/429367}

\bibitem[{{Lin} \& {Papaloizou}(1986)}]{Lin_1986}
{Lin}, D.~N.~C., \& {Papaloizou}, J. 1986, \apj, 309, 846, \dodoi{10.1086/164653}

\bibitem[{Lin \& Pierens(2018)}]{Lin_2018}
Lin, M.-K., \& Pierens, A. 2018, Monthly Notices of the Royal Astronomical Society, 478, 575–591, \dodoi{10.1093/mnras/sty947}

\bibitem[{{Lodato} {et~al.}(2019){Lodato}, {Dipierro}, {Ragusa}, {Long}, {Herczeg}, {Pascucci}, {Pinilla}, {Manara}, {Tazzari}, {Liu}, {Mulders}, {Harsono}, {Boehler}, {M{\'e}nard}, {Johnstone}, {Salyk}, {van der Plas}, {Cabrit}, {Edwards}, {Fischer}, {Hendler}, {Nisini}, {Rigliaco}, {Avenhaus}, {Banzatti}, \& {Gully-Santiago}}]{Lodato_2019}
{Lodato}, G., {Dipierro}, G., {Ragusa}, E., {et~al.} 2019, \mnras, 486, 453, \dodoi{10.1093/mnras/stz913}

\bibitem[{Long {et~al.}(2018)Long, Pinilla, Herczeg, Harsono, Dipierro, Pascucci, Hendler, Tazzari, Ragusa, Salyk, Edwards, Lodato, van~de Plas, Johnstone, Liu, Boehler, Cabrit, Manara, Menard, Mulders, Nisini, Fischer, Rigliaco, Banzatti, Avenhaus, \& Gully-Santiago}]{Long_2018}
Long, F., Pinilla, P., Herczeg, G.~J., {et~al.} 2018, The Astrophysical Journal, 869, 17, \dodoi{10.3847/1538-4357/aae8e1}

\bibitem[{{Lovelace} {et~al.}(1999){Lovelace}, {Li}, {Colgate}, \& {Nelson}}]{Lovelace_1999}
{Lovelace}, R.~V.~E., {Li}, H., {Colgate}, S.~A., \& {Nelson}, A.~F. 1999, \apj, 513, 805, \dodoi{10.1086/306900}

\bibitem[{{Masset}(2000)}]{Masset_2000}
{Masset}, F. 2000, \aaps, 141, 165, \dodoi{10.1051/aas:2000116}

\bibitem[{{Meheut}(2013{\natexlab{a}})}]{Meheut_2013b}
{Meheut}, H. 2013{\natexlab{a}}, in European Physical Journal Web of Conferences, Vol.~46, European Physical Journal Web of Conferences, 03001, \dodoi{10.1051/epjconf/20134603001}

\bibitem[{{Meheut}(2013{\natexlab{b}})}]{Meheut_2013c}
{Meheut}, H. 2013{\natexlab{b}}, in Poster presented at Protostars and Planets VI

\bibitem[{Meheut {et~al.}(2010)Meheut, Casse, Varniere, \& Tagger}]{Meheut_2010}
Meheut, H., Casse, F., Varniere, P., \& Tagger, M. 2010, Astronomy and Astrophysics, 516, A31, \dodoi{10.1051/0004-6361/201014000}

\bibitem[{{Meheut} {et~al.}(2012{\natexlab{a}}){Meheut}, {Keppens}, {Casse}, \& {Benz}}]{Meheut_2012b}
{Meheut}, H., {Keppens}, R., {Casse}, F., \& {Benz}, W. 2012{\natexlab{a}}, \aap, 542, A9, \dodoi{10.1051/0004-6361/201118500}

\bibitem[{{Meheut} {et~al.}(2013){Meheut}, {Lovelace}, \& {Lai}}]{Meheut_2013a}
{Meheut}, H., {Lovelace}, R.~V.~E., \& {Lai}, D. 2013, \mnras, 430, 1988, \dodoi{10.1093/mnras/stt022}

\bibitem[{{Meheut} {et~al.}(2012{\natexlab{b}}){Meheut}, {Meliani}, {Varniere}, \& {Benz}}]{Meheut_2012c}
{Meheut}, H., {Meliani}, Z., {Varniere}, P., \& {Benz}, W. 2012{\natexlab{b}}, \aap, 545, A134, \dodoi{10.1051/0004-6361/201219794}

\bibitem[{{Meheut} {et~al.}(2012{\natexlab{c}}){Meheut}, {Yu}, \& {Lai}}]{Meheut_2012a}
{Meheut}, H., {Yu}, C., \& {Lai}, D. 2012{\natexlab{c}}, \mnras, 422, 2399, \dodoi{10.1111/j.1365-2966.2012.20789.x}

\bibitem[{Miranda {et~al.}(2017)Miranda, Li, Li, \& Jin}]{Miranda_2017}
Miranda, R., Li, H., Li, S., \& Jin, S. 2017, The Astrophysical Journal, 835, 118, \dodoi{10.3847/1538-4357/835/2/118}

\bibitem[{Miranda \& Rafikov(2019)}]{Miranda_2019}
Miranda, R., \& Rafikov, R.~R. 2019, The Astrophysical Journal Letters, 878, L9, \dodoi{10.3847/2041-8213/ab22a7}

\bibitem[{{Miranda} \& {Rafikov}(2020)}]{Miranda_2020}
{Miranda}, R., \& {Rafikov}, R.~R. 2020, \apj, 892, 65, \dodoi{10.3847/1538-4357/ab791a}

\bibitem[{{Muto} {et~al.}(2010){Muto}, {Suzuki}, \& {Inutsuka}}]{Muto_2010}
{Muto}, T., {Suzuki}, T.~K., \& {Inutsuka}, S.-i. 2010, \apj, 724, 448, \dodoi{10.1088/0004-637X/724/1/448}

\bibitem[{{Narayan} {et~al.}(1987){Narayan}, {Goldreich}, \& {Goodman}}]{Narayan_1987}
{Narayan}, R., {Goldreich}, P., \& {Goodman}, J. 1987, \mnras, 228, 1, \dodoi{10.1093/mnras/228.1.1}

\bibitem[{Ono {et~al.}(2016)Ono, Muto, Takeuchi, \& Nomura}]{Ono_2016}
Ono, T., Muto, T., Takeuchi, T., \& Nomura, H. 2016, The Astrophysical Journal, 823, 84, \dodoi{10.3847/0004-637x/823/2/84}

\bibitem[{Ono {et~al.}(2018)Ono, Muto, Tomida, \& Zhu}]{Ono_2018}
Ono, T., Muto, T., Tomida, K., \& Zhu, Z. 2018, The Astrophysical Journal, 864, 70, \dodoi{10.3847/1538-4357/aad54d}

\bibitem[{Paardekooper {et~al.}(2010)Paardekooper, Lesur, \& Papaloizou}]{Paardekooper_2010}
Paardekooper, S.-J., Lesur, G., \& Papaloizou, J. C.~B. 2010, The Astrophysical Journal, 725, 146–158, \dodoi{10.1088/0004-637x/725/1/146}

\bibitem[{{P{\'e}rez} {et~al.}(2018){P{\'e}rez}, {Benisty}, {Andrews}, {Isella}, {Dullemond}, {Huang}, {Kurtovic}, {Guzm{\'a}n}, {Zhu}, {Birnstiel}, {Zhang}, {Carpenter}, {Wilner}, {Ricci}, {Bai}, {Weaver}, \& {{\"O}berg}}]{Perez_2018}
{P{\'e}rez}, L.~M., {Benisty}, M., {Andrews}, S.~M., {et~al.} 2018, \apjl, 869, L50, \dodoi{10.3847/2041-8213/aaf745}

\bibitem[{Pierens \& Lin(2018)}]{Pierens_2018}
Pierens, A., \& Lin, M.-K. 2018, Monthly Notices of the Royal Astronomical Society, \dodoi{10.1093/mnras/sty1314}

\bibitem[{Pinilla {et~al.}(2012)Pinilla, Benisty, \& Birnstiel}]{Pinilla_2012}
Pinilla, P., Benisty, M., \& Birnstiel, T. 2012, Astronomy and amp; Astrophysics, 545, A81, \dodoi{10.1051/0004-6361/201219315}

\bibitem[{{Pinte} {et~al.}(2023){Pinte}, {Teague}, {Flaherty}, {Hall}, {Facchini}, \& {Casassus}}]{Pinte_2023}
{Pinte}, C., {Teague}, R., {Flaherty}, K., {et~al.} 2023, in Astronomical Society of the Pacific Conference Series, Vol. 534, Protostars and Planets VII, ed. S.~{Inutsuka}, Y.~{Aikawa}, T.~{Muto}, K.~{Tomida}, \& M.~{Tamura}, 645, \dodoi{10.48550/arXiv.2203.09528}

\bibitem[{{Richard} {et~al.}(2013){Richard}, {Barge}, \& {Le Diz{\`e}s}}]{Richard_2013}
{Richard}, S., {Barge}, P., \& {Le Diz{\`e}s}, S. 2013, \aap, 559, A30, \dodoi{10.1051/0004-6361/201322175}

\bibitem[{{Segura-Cox} {et~al.}(2020){Segura-Cox}, {Schmiedeke}, {Pineda}, {Stephens}, {Fern{\'a}ndez-L{\'o}pez}, {Looney}, {Caselli}, {Li}, {Mundy}, {Kwon}, \& {Harris}}]{Segura-Cox_2020}
{Segura-Cox}, D.~M., {Schmiedeke}, A., {Pineda}, J.~E., {et~al.} 2020, \nat, 586, 228, \dodoi{10.1038/s41586-020-2779-6}

\bibitem[{{Shakura} \& {Sunyaev}(1973)}]{Shakura_1973}
{Shakura}, N.~I., \& {Sunyaev}, R.~A. 1973, \aap, 24, 337

\bibitem[{Sierra {et~al.}(2019)Sierra, Lizano, Macías, Carrasco-González, Osorio, \& Flock}]{Sierra_2019}
Sierra, A., Lizano, S., Macías, E., {et~al.} 2019, The Astrophysical Journal, 876, 7, \dodoi{10.3847/1538-4357/ab1265}

\bibitem[{{Stone} \& {Gardiner}(2010)}]{Stone_2010}
{Stone}, J.~M., \& {Gardiner}, T.~A. 2010, \apjs, 189, 142, \dodoi{10.1088/0067-0049/189/1/142}

\bibitem[{Stone {et~al.}(2020)Stone, Tomida, White, \& Felker}]{Stone_2020}
Stone, J.~M., Tomida, K., White, C.~J., \& Felker, K.~G. 2020, The Astrophysical Journal Supplement Series, 249, 4, \dodoi{10.3847/1538-4365/ab929b}

\bibitem[{{The CASA Team} {et~al.}(2022){The CASA Team}, Bean, Bhatnagar, Castro, Meyer, Emonts, Garcia, Garwood, Golap, Villalba, Harris, Hayashi, Hoskins, Hsieh, Jagannathan, Kawasaki, Keimpema, Kettenis, Lopez, Marvil, Masters, McNichols, Mehringer, Miel, Moellenbrock, Montesino, Nakazato, Ott, Petry, Pokorny, Raba, Rau, Schiebel, Schweighart, Sekhar, Shimada, Small, Steeb, Sugimoto, Suoranta, Tsutsumi, van Bemmel, Verkouter, Wells, Xiong, Szomoru, Griffith, Glendenning, \& Kern}]{The_CASA_Team_2022}
{The CASA Team}, Bean, B., Bhatnagar, S., {et~al.} 2022, Publications of the Astronomical Society of the Pacific, 134, 114501, \dodoi{10.1088/1538-3873/ac9642}

\bibitem[{{van der Marel} {et~al.}(2016){van der Marel}, {van Dishoeck}, {Bruderer}, {Andrews}, {Pontoppidan}, {Herczeg}, {van Kempen}, \& {Miotello}}]{van_der_Marel_2016}
{van der Marel}, N., {van Dishoeck}, E.~F., {Bruderer}, S., {et~al.} 2016, \aap, 585, A58, \dodoi{10.1051/0004-6361/201526988}

\bibitem[{van~der Marel {et~al.}(2013)van~der Marel, van Dishoeck, Bruderer, Birnstiel, Pinilla, Dullemond, van Kempen, Schmalzl, Brown, Herczeg, Mathews, \& Geers}]{van_der_Marel_2013}
van~der Marel, N., van Dishoeck, E.~F., Bruderer, S., {et~al.} 2013, Science, 340, 1199–1202, \dodoi{10.1126/science.1236770}

\bibitem[{van~der Marel {et~al.}(2020)van~der Marel, Birnstiel, Garufi, Ragusa, Christiaens, Price, Sallum, Muley, Francis, \& Dong}]{van_der_Marel_2020}
van~der Marel, N., Birnstiel, T., Garufi, A., {et~al.} 2020, The Astronomical Journal, 161, 33, \dodoi{10.3847/1538-3881/abc3ba}

\bibitem[{Zhang {et~al.}(2023)Zhang, Huang, \& Dong}]{Zhang_2023}
Zhang, M., Huang, P., \& Dong, R. 2023, The dependence of the structure of planet-opened gaps in protoplanetary disks on radiative cooling.
\newblock \doarXiv{2310.11757}

\bibitem[{{Zhang} {et~al.}(2018){Zhang}, {Zhu}, {Huang}, {Guzm{\'a}n}, {Andrews}, {Birnstiel}, {Dullemond}, {Carpenter}, {Isella}, {P{\'e}rez}, {Benisty}, {Wilner}, {Baruteau}, {Bai}, \& {Ricci}}]{Zhang_2018}
{Zhang}, S., {Zhu}, Z., {Huang}, J., {et~al.} 2018, \apjl, 869, L47, \dodoi{10.3847/2041-8213/aaf744}

\bibitem[{{Zhu} {et~al.}(2014){Zhu}, {Stone}, {Rafikov}, \& {Bai}}]{Zhu_2014}
{Zhu}, Z., {Stone}, J.~M., {Rafikov}, R.~R., \& {Bai}, X.-n. 2014, \apj, 785, 122, \dodoi{10.1088/0004-637X/785/2/122}

\end{thebibliography}
\end{CJK*}
\end{document}